\documentclass[10pt, twocolumn, aps, nofootinbib, longbibliography, nobibnotes]{revtex4-1}
\pdfoutput=1

\usepackage{amsfonts, amssymb, amsmath, amsthm}
\usepackage{latexsym}
\usepackage[tracking=smallcaps]{microtype}	
\usepackage{url}
\usepackage{color}
\definecolor{DarkGray}{rgb}{0.1,0.1,0.5}
\usepackage[colorlinks=true,breaklinks, linkcolor=DarkGray,citecolor=DarkGray,urlcolor=DarkGray]{hyperref}	

\usepackage{graphicx}
\usepackage[tight, TABBOTCAP]{subfigure}

\def\S {{\mathcal S}}

\newcommand{\identity}{\ensuremath{\boldsymbol{1}}} 

\newcounter{sprows}

\newlength{\spheight}
\newlength{\spraise}

\newcommand{\comment}[1]{\emph{\color{blue}Comment:\color{black} #1}} 
\newlength{\commentslength}
\newcommand{\comments}[1]{
\hspace{-2\parindent}
\addtolength{\commentslength}{-\commentslength}
\addtolength{\commentslength}{\linewidth}
\addtolength{\commentslength}{-\parindent}
\fcolorbox{blue}{white}{\smallskip\begin{minipage}[c]{\commentslength}
\emph{Comments:}\begin{itemize}#1\end{itemize}\end{minipage}}\bigskip
}
\newcommand{\rem}[1]{}

\newtheorem{theorem}{Theorem}

\newtheorem{claim}[theorem]{Claim}

\newfont{\subsubsecfnt}{ptmri8t at 11pt}
\renewcommand{\subparagraph}[1]{\smallskip{\subsubsecfnt #1.}}

\newcommand{\eqnref}[1]{\hyperref[#1]{{(\ref*{#1})}}}
\newcommand{\thmref}[1]{\hyperref[#1]{{Theorem~\ref*{#1}}}}
\newcommand{\lemref}[1]{\hyperref[#1]{{Lemma~\ref*{#1}}}}
\newcommand{\corref}[1]{\hyperref[#1]{{Corollary~\ref*{#1}}}}
\newcommand{\defref}[1]{\hyperref[#1]{{Definition~\ref*{#1}}}}
\newcommand{\secref}[1]{\hyperref[#1]{{Sec.~\ref*{#1}}}}
\newcommand{\figref}[1]{\hyperref[#1]{{Fig.~\ref*{#1}}}}
\newcommand{\tabref}[1]{\hyperref[#1]{{Table~\ref*{#1}}}}
\newcommand{\remref}[1]{\hyperref[#1]{{Remark~\ref*{#1}}}}
\newcommand{\appref}[1]{\hyperref[#1]{{Appendix~\ref*{#1}}}}
\newcommand{\claimref}[1]{\hyperref[#1]{{Claim~\ref*{#1}}}}
\newcommand{\factref}[1]{\hyperref[#1]{{Fact~\ref*{#1}}}}
\newcommand{\propref}[1]{\hyperref[#1]{{Proposition~\ref*{#1}}}}
\newcommand{\exampleref}[1]{\hyperref[#1]{{Example~\ref*{#1}}}}
\newcommand{\conjref}[1]{\hyperref[#1]{{Conjecture~\ref*{#1}}}}

\allowdisplaybreaks[1]

\def\COLOR{}
\ifdefined\COLOR

\else

\fi

\definecolor{Cayenne}{rgb}{0.5,0,0}
\definecolor{Midnight}{rgb}{0,0,0.5}
\definecolor{Plum}{rgb}{0.5,0,0.5}
\definecolor{Teal}{rgb}{0,0.5,0.5}
\definecolor{Clover}{rgb}{0,0.5,0}
\definecolor{Maroon}{rgb}{0.5,0,0.25}
\definecolor{Ocean}{rgb}{0,0.25,0.5}
\definecolor{Tangerine}{rgb}{1,0.5,0}
\definecolor{Strawberry}{rgb}{1,0,0.5}
\definecolor{Fern}{rgb}{0.25,0.5,0}
\definecolor{Aqua}{rgb}{0,0.5,1}
\definecolor{Moss}{rgb}{0,0.5,0.25}
\definecolor{Mocha}{rgb}{0.5,0.25,0}
\definecolor{Lemon}{rgb}{1,1,0}
\definecolor{Asparagus}{rgb}{0.5,0.5,0}
\definecolor{Grape}{rgb}{0.5,0,1}
\definecolor{Iron}{rgb}{.3,.3,.3}
\definecolor{Steel}{rgb}{.4,.4,.4}

\usepackage{enumitem} 

\makeatletter
\let\save@mathaccent\mathaccent
\newcommand*\if@single[3]{%
  \setbox0\hbox{${\mathaccent"0362{#1}}^H$}%
  \setbox2\hbox{${\mathaccent"0362{\kern0pt#1}}^H$}%
  \ifdim\ht0=\ht2 #3\else #2\fi
  }
\newcommand*\rel@kern[1]{\kern#1\dimexpr\macc@kerna}
\newcommand*\widebar[1]{\@ifnextchar^{{\wide@bar{#1}{0}}}{\wide@bar{#1}{1}}}
\newcommand*\wide@bar[2]{\if@single{#1}{\wide@bar@{#1}{#2}{1}}{\wide@bar@{#1}{#2}{2}}}
\newcommand*\wide@bar@[3]{%
  \begingroup
  \def\mathaccent##1##2{%
    \let\mathaccent\save@mathaccent
    \if#32 \let\macc@nucleus\first@char \fi
    \setbox\z@\hbox{$\macc@style{\macc@nucleus}_{}$}%
    \setbox\tw@\hbox{$\macc@style{\macc@nucleus}{}_{}$}%
    \dimen@\wd\tw@
    \advance\dimen@-\wd\z@
    \divide\dimen@ 3
    \@tempdima\wd\tw@
    \advance\@tempdima-\scriptspace
    \divide\@tempdima 10
    \advance\dimen@-\@tempdima
    \ifdim\dimen@>\z@ \dimen@0pt\fi
    \rel@kern{0.6}\kern-\dimen@
    \if#31
      \overline{\rel@kern{-0.6}\kern\dimen@\macc@nucleus\rel@kern{0.4}\kern\dimen@}%
      \advance\dimen@0.4\dimexpr\macc@kerna
      \let\final@kern#2%
      \ifdim\dimen@<\z@ \let\final@kern1\fi
      \if\final@kern1 \kern-\dimen@\fi
    \else
      \overline{\rel@kern{-0.6}\kern\dimen@#1}%
    \fi
  }%
  \macc@depth\@ne
  \let\math@bgroup\@empty \let\math@egroup\macc@set@skewchar
  \mathsurround\z@ \frozen@everymath{\mathgroup\macc@group\relax}%
  \macc@set@skewchar\relax
  \let\mathaccentV\macc@nested@a
  \if#31
    \macc@nested@a\relax111{#1}%
  \else
    \def\gobble@till@marker##1\endmarker{}%
    \futurelet\first@char\gobble@till@marker#1\endmarker
    \ifcat\noexpand\first@char A\else
      \def\first@char{}%
    \fi
    \macc@nested@a\relax111{\first@char}%
  \fi
  \endgroup
}
\makeatother

\def\llbracket{{[\![}}
\def\rrbracket{{]\!]}}

\renewcommand{\comment}[1]{}
\renewcommand{\comments}[1]{}

\begin{document}
\def\compilefullpaper{}

\title{Fault-tolerant quantum error correction \\ for Steane's seven-qubit color code with few or no extra qubits}
\author{Ben W. Reichardt}
\affiliation{University of Southern California}

\begin{abstract}
Steane's seven-qubit quantum code is a natural choice for fault-tolerance experiments because it is small and just two extra qubits are enough to correct errors.  However, the two-qubit error-correction technique, known as ``flagged" syndrome extraction, works slowly, measuring only one syndrome at a time.  This is a disadvantage in experiments with high qubit rest error rates.  We extend the technique to extract multiple syndromes at once, without needing more qubits.  Qubits for different syndromes can flag errors in each other.  This gives equally fast and more qubit-efficient alternatives to Steane's error-correction method, and also conforms to planar geometry constraints.  

We further show that Steane's code and some others can be error-corrected with no extra qubits, provided there are at least two code blocks.  The rough idea is that two seven-qubit codewords can be temporarily joined into a twelve-qubit code, freeing two qubits for flagged syndrome measurement.  
\end{abstract}

\maketitle

\section{Introduction}

\begin{figure}[b]
\centering
\begin{equation*}
\raisebox{-1.7cm}{\includegraphics[scale=1]{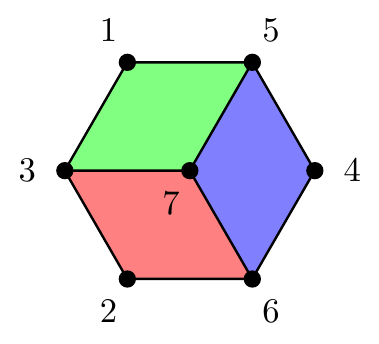}}
\begin{array}{r c c c c c c c}
&\scalebox{.8}{1}&\scalebox{.8}{2}&\scalebox{.8}{3}&\scalebox{.8}{4}&\scalebox{.8}{5}&\scalebox{.8}{6}&\scalebox{.8}{7}\\
&I&I&I&X&X&X&X\\
&I&X&X&I&I&X&X\\
&X&I&X&I&X&I&X\\
&I&I&I&Z&Z&Z&Z\\
&I&Z&Z&I&I&Z&Z\\
&Z&I&Z&I&Z&I&Z\\
\cline{2-8}
\widebar X = &X&X&X&X&X&X&X\\
\widebar Z = &Z&Z&Z&Z&Z&Z&Z
\end{array}
\end{equation*}
\caption{Stabilizers for the Steane color code.  For each \mbox{plaquette} on the left there are $X$ and $Z$ stabilizers supported on the incident vertices.  Logical $X$ and $Z$ are transversal.  
} \label{f:steanecolor}
\end{figure}

\begin{figure}[b]
\centering
\begin{equation*}
\raisebox{.34cm}{\includegraphics[scale=.769]{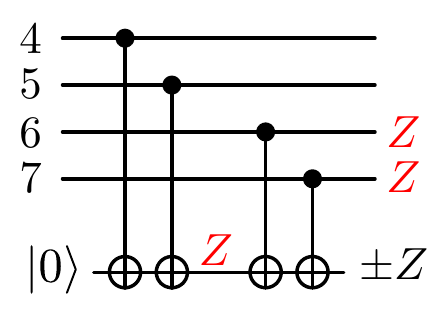}}
\qquad 
\includegraphics[scale=.769]{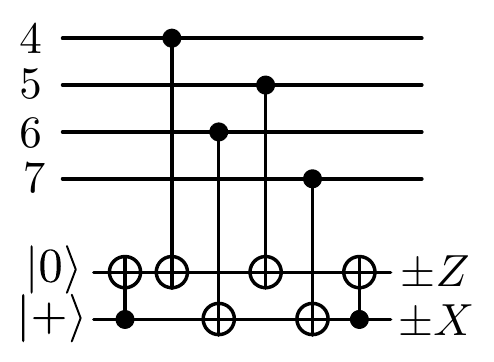}
\end{equation*}
\caption{
Flagged syndrome extraction.  Briefly, the first circuit measures into one extra qubit the syndrome of the $IIIZZZZ$ stabilizer, to detect $X$ or $Y$ errors on the last four qubits.  However, a $Z$ fault halfway through spreads to a $Z_6 Z_7$ error on the data, and this error, equivalent to $Z_1 \protect\widebar Z$, is uncorrectable.  
The right circuit adds a second, ``flag" qubit, which detects bad $Z$ faults.  If the $X$ measurement is nontrivial, then all $X$ syndromes are measured and the best correction is applied, e.g., if the syndromes indicate a $Z_1$ error then the correction is actually $Z_6 Z_7$.  For details, see~\protect{\cite{ChaoReichardt17errorcorrection}}.  
} \label{f:flagerrorcorrection}
\end{figure}

Steane's $\llbracket 7,1,3 \rrbracket$ quantum code~\cite{Steane96css} is a color code~\cite{BombinMartindelgado06colorcode} on seven qubits; see \figref{f:steanecolor}.  The code protects one logical qubit to distance three, so it can correct any one-qubit error.  However, extra qubits are needed for fault-tolerant error correction.  Steane's method~\cite{Steane97, Steane02} uses at least a full code block of seven extra qubits.  Shor-style error correction~\cite{Shor96} requires either five or four~\cite{DiVincenzoAliferis06slow} extra qubits.  
The error-correction overhead was reduced to two extra qubits by Yoder and Kim~\cite{YoderKim16trianglecodes}, by a method of ``flagged" syndrome extraction~\cite{ChaoReichardt17errorcorrection, ChamberlandBeverland17flags, TansuwannontChamberlandLeung18flag}; see \figref{f:flagerrorcorrection}.  

The two-qubit flagged syndrome-extraction technique works for many codes, and is promising for testing fault-tolerant error correction on small quantum computing devices.  But can we do better?  There are two natural avenues for improvement.  

\renewcommand{\theparagraph}{\arabic{paragraph}}

\begin{figure}[b]
\centering
\begin{equation*}
\raisebox{-1.3cm}{\includegraphics[scale=1]{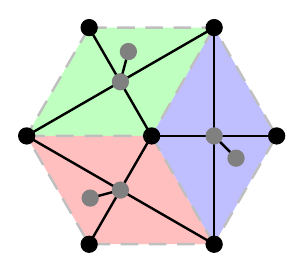}}
\qquad
\raisebox{-1.3cm}{\includegraphics[scale=1]{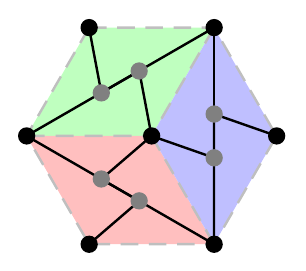}}
\end{equation*}
\caption{Layouts for $\llbracket 7,1,3 \rrbracket$ flag error correction with two extra qubits per plaquette, $13$ total.  The advantage of the right layout, from~\cite{YoderKim16trianglecodes}, is that qubits have degree only three.  
} \label{f:steane2d6ancillas}
\end{figure}

\smallskip
\paragraph{Gate locality and memory faults.}  
In an ion trap quantum computing experiment, gates can be applied between arbitrary qubits within the same trap; and with multiple traps, qubits can be moved between them.  One can envision using the same two extra qubits to measure each code stabilizer.  
However, in a superconducting qubit experiment, for example, qubits are placed at fixed locations in a 2D plane and only some two-qubit interactions are possible.  It is challenging to place nine qubits with all the necessary two-qubit couplings for $\llbracket 7,1,3 \rrbracket$ code fault-tolerant error correction.  
Yoder and Kim~\cite{YoderKim16trianglecodes} suggest placing two extra qubits in each plaquette, with the interactions shown in \figref{f:steane2d6ancillas}.  This uses $13$ qubits total.  

\begin{figure}
\centering
\raisebox{-1.25cm}{\includegraphics[scale=1]{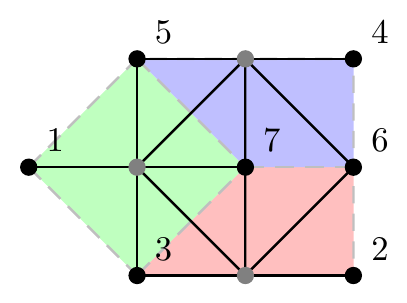}}
\caption{
A grid layout for the Steane code with one extra qubit per plaquette, ten qubits total.  
} \label{f:steane2d3ancillas}
\end{figure}

We study syndrome extraction with one extra qubit per plaquette---$10$ qubits total---as in \figref{f:steane2d3ancillas}.  In addition to using fewer qubits, we address another problem of superconducting qubit systems relative to ion traps: higher idle qubit error rates.  If qubits accumulate errors rapidly even at rest, then it is important to extract multiple syndromes in parallel instead of measuring them one at a time, especially when qubit initialization or measurement is slow.  (For example, the measurement time in~\cite{Barendsetal14superconducting} is over four times the CZ gate time.)  We show how three of the six syndromes can be measured together, with the layout of \figref{f:steane2d3ancillas}, so all syndromes can be measured in two measurement rounds.  

Our technique extends the flagged syndrome-extraction method.  In \figref{f:flagerrorcorrection}, one extra qubit stores the syndrome and the other extra qubit is used to flag possible correlated errors.  Instead, two syndromes can be extracted at once, each into an extra qubit; and by coupling the extra qubits appropriately, one serves to flag correlated errors from the other.  Neither extra qubit is a dedicated flag qubit; instead the syndrome qubits reciprocally flag each other.  

\begin{figure}
\centering
\raisebox{0cm}{\includegraphics[scale=.769]{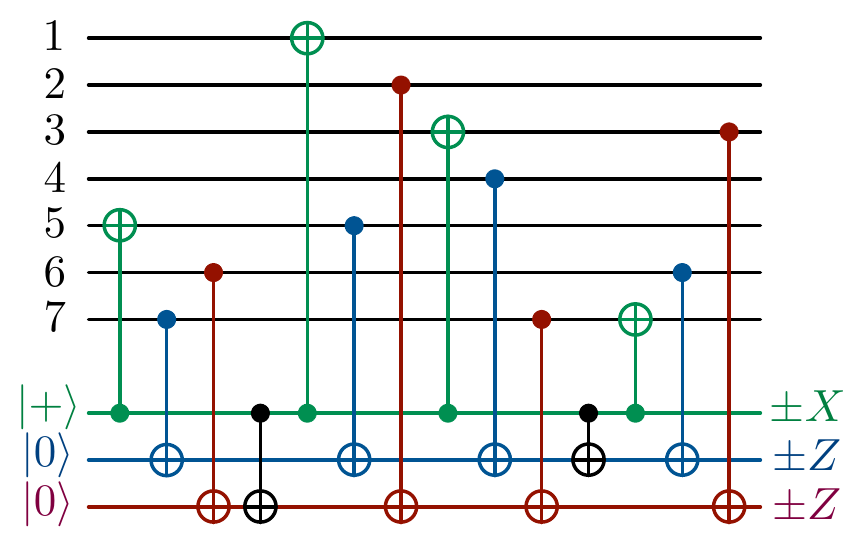}}
\caption{
This circuit simultaneously extracts the syndromes for $X_{1,3,5,7}$ (green gates), $Z_{4,5,6,7}$ (blue) and $Z_{2,3,6,7}$ (red).  The gates are local with the geometry of \figref{f:steane2d3ancillas}.  
The $X$ syndrome measurement catches correlated $Z$ errors that can be created from the $Z$ syndrome measurements, while the first $Z$ syndrome measurement catches the correlated error $X_{3,7}$ that can be created from the $X$ syndrome measurement.  See \secref{s:steaneparallel3syndromes} for details.  
} \label{f:steaneparallel3syndromesunlabeled}
\end{figure}

This idea extends to extracting three syndromes at once, using the circuit of \figref{f:steaneparallel3syndromesunlabeled}.  
It also extends to other codes, such as the $\llbracket 4,2,2 \rrbracket$ and $\llbracket 12,2,3 \rrbracket$ color codes.  For the $\llbracket 5,1,3 \rrbracket$ code and the $\llbracket 15,7,3 \rrbracket$ Hamming code, one can extract two syndromes in parallel with three extra qubits---the third, flag qubit being shared between the two syndrome measurements (see \secref{s:sharedflag}).  

\smallskip
\paragraph{Error correction without extra qubits.}
Are two extra qubits necessary for fault-tolerant error correction?  Not always.  One extra qubit is enough for some codes, such as the Bacon-Shor $\llbracket 9,1,3 \rrbracket$ code~\cite{AliferisCross06BaconShorft, LiMillerBrown18baconshor}.  We study color codes requiring only one extra qubit in Secs.~\ref{s:onequbitcolorcode} and~\ref{s:mergedcolorcodes}.  
With Steane's code, though, one cannot safely measure a syndrome into just a single extra qubit because the measurement can create undetected weight-two $X$ or $Z$ errors; and as the code is a perfect CSS code, weight-two $X$ or $Z$ errors are all uncorrectable.  

\begin{figure}
\centering
\raisebox{0cm}{\includegraphics[scale=.192]{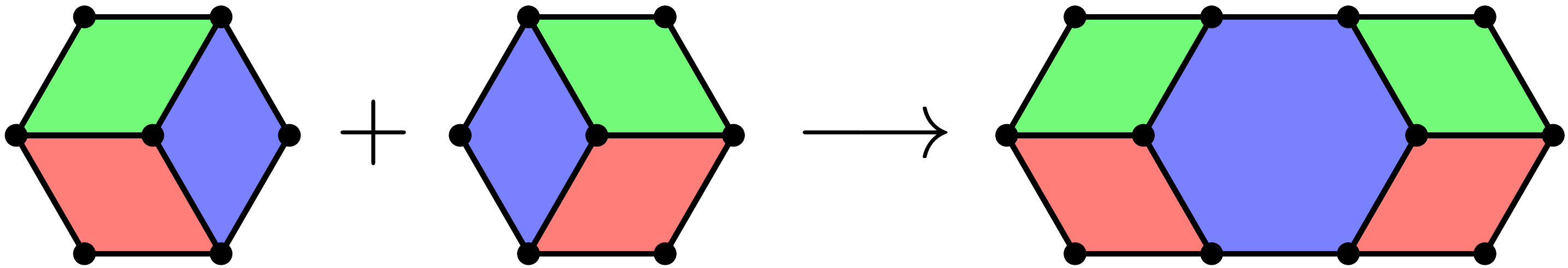}} 
\caption{Two $\llbracket 7,1,3 \rrbracket$ Steane color codes can be merged into a $\llbracket 12,2,3 \rrbracket$ code.  This frees two qubits for error correction.  
} \label{f:steanecodemerge1223}
\end{figure}

Surprisingly, we show that zero extra qubits are needed for fault-tolerant error correction of Steane's code.  The catch is that there must be at least two code blocks, $14$ qubits total.  Intuitively, two seven-qubit code blocks can be merged into a $\llbracket 12,2,3 \rrbracket$ color code, as shown in \figref{f:steanecodemerge1223}, and this frees two qubits for flagged error correction.  Our actual construction, in \secref{s:steanesteaneerrorcorrection}, is a bit simpler, using one code block to flag errors in the other without converting to the $\llbracket 12,2,3 \rrbracket$ color code.  The method also works for erasure codes and for the $\llbracket 15,7,3 \rrbracket$ Hamming code (\secref{s:codecodeerrorcorrection}), but it does use more CNOT gates than two-qubit flagged error correction.  

\section{Parallel syndrome extraction}

Measuring the code syndromes one at a time might require only a few extra qubits, but is also slow.  This is bad if the rest error rate is high, especially when qubit initializations or measurements are slow.  For a CSS code, Steane-style error correction~\cite{Steane97, Steane02} works faster, extracting half the syndromes at once---but it requires at least a full code block of extra qubits.  Knill's scheme~\cite{Knill03erasure} extracts all the syndromes at once, and works even for non-CSS codes, but requires at least two code blocks of extra qubits.  

In this section, we study intermediate syndrome-extraction methods, that extract two or three syndromes together and yet are also qubit-efficient.  For some codes, such as the Steane code, multiple syndromes can be measured at once with only one extra qubit per syndrome, with one syndrome serving also as a flag to catch correlated errors created while measuring another syndrome (Sec.~\ref{s:parallelsyndromesteanecode} and~\ref{s:parallelsyndromeerasurecode}).  For other codes, such as the $\llbracket 5,1,3 \rrbracket$ or $\llbracket 15,7,3 \rrbracket$ Hamming codes, one flag qubit can be shared with multiple syndrome measurements (\secref{s:sharedflag}).  For the $\llbracket 15,7,3 \rrbracket$ code, extracting two syndromes at once not only works faster than measuring them sequentially, but can also use fewer gates, e.g., $17$ instead of $20$ CNOT~gates.

\subsection{Steane code parallel syndrome extraction} \label{s:parallelsyndromesteanecode}

Syndrome extraction for the $\llbracket 7,1,3 \rrbracket$ Steane code can be parallelized, with one syndrome acting as flag for another.

\subsubsection{Two syndromes in parallel, $X_{4,5,6,7}$ and $Z_{2,3,6,7}$}

Consider the following circuit: 
\begin{equation} \label{e:steaneparallel2syndromes}
\raisebox{-2.1cm}{\includegraphics[scale=.769]{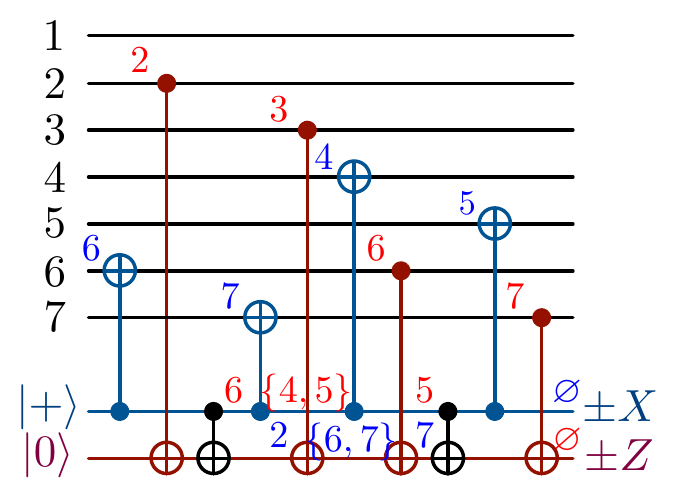}}
\end{equation}
The blue CNOT gates extract the $X_{4,5,6,7}$ syndrome, while the red gates extract the $Z_{2,3,6,7}$ syndrome.  The black gates formally commute with the other gates and cancel.  

Error correction works by extracting two syndromes at a time using three circuits like the one above.  If a nontrivial syndrome is measured, then all six syndromes are measured and an appropriate correction applied.  

For a distance-three, perfect CSS code, the Aliferis-Gottesman-Preskill (AGP) extended rectangle conditions for fault-tolerant error correction~\cite{AliferisGottesmanPreskill05} simplify to: 
\begin{enumerate}[leftmargin=*]
\item 
If the input error's $X$ and $Z$ components both have weight $\leq 1$ and no faults occur during error correction, then the state is mapped to the closest codeword.  
\item 
If the input is perfect and at most one location fault occurs during error correction, then the output error has $X$ and $Z$ components both of weight $\leq 1$.  
\end{enumerate}

For analyzing fault tolerance, one can consider $X$ and $Z$ errors separately.  It is enough to show that if the input to error correction is perfect and there is at most one $X$ fault, then the output $X$ error has weight $\leq 1$, and similarly for $Z$ errors.  

The main concern here is that a single $X$ failure at the location marked ${\color{red}\{4,5\}}$ will spread via the subsequent CNOT gates into a weight-two data error, $X_{4,5} \sim X_1 \widebar X$.  And a $Z$ failure at location ${\color{blue}\{6,7\}}$ spreads to $Z_{6,7} \sim Z_1 \widebar Z$.  We do not want a single failure to cause a logical error.  The black gates are there to catch such bad faults.  

Assuming at most one failure, there are two cases.  First, if the syndromes are both trivial, then the data gets at most a weight-one error.  (The only bad failure locations, marked ${\color{red}\{4,5\}}$ and ${\color{blue}\{6,7\}}$, are both caught.)  
Second, if a syndrome is nontrivial, then all possible data errors are distinguishable.  Indeed, in Eq.~\eqnref{e:steaneparallel2syndromes} we have marked in red all inequivalent $X$ failure locations that can cause a nontrivial $Z$ syndrome measurement, and in blue all $Z$ failure locations that can cause a nontrivial $X$ syndrome measurement.  The possible $X$ errors are then $\identity, X_2, X_3, X_5, X_6, X_7, X_{4,5}$, and the possible $Z$ errors are $\identity, Z_2, Z_4, Z_5, Z_6, Z_7, Z_{6,7}$.  As the errors are distinguishable, they can be corrected after measuring the six syndromes.  (In particular, notice that if the $Z$ measurement in Eq.~\eqnref{e:steaneparallel2syndromes} is nontrivial, then the correction for the $+,+,-$ $Z$ syndrome is $X_{4,5}$, and not $X_1$.)

\subsubsection{Two dual syndromes in parallel, $X_{4,5,6,7}$ and $Z_{4,5,6,7}$}

This circuit similarly extracts two syndromes in parallel, using one syndrome to flag the other: 
\begin{equation*}
\raisebox{0cm}{\includegraphics[scale=.769]{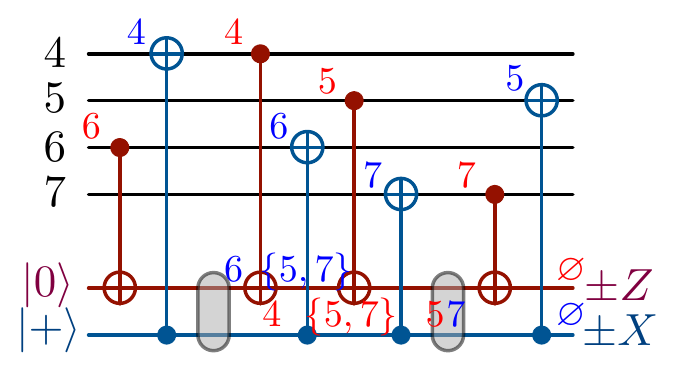}}
\end{equation*}
In this case, the measured syndromes are for the dual stabilizers $X_{4,5,6,7}$ and $Z_{4,5,6,7}$.  The circuit uses two extra qubits for the plaquette.  The CNOT gates are geometrically local according to the interaction geometry on the right-hand side of \figref{f:steane2d6ancillas}, provided that the extra qubits are switched at the two shaded marks.  (A single failure at a swap location can create errors on both involved qubits.)  Note that in this circuit, unlike in Eq.~\eqnref{e:steaneparallel2syndromes}, there is no CNOT gate between the two extra qubits; nonetheless, errors on one can spread to the other via the carefully ordered CNOT gates.  

As in Eq.~\eqnref{e:steaneparallel2syndromes}, we have marked in red $X$ fault locations that trigger the $Z$ measurement,
and in blue $Z$ fault locations that trigger the $X$ measurement.  (For example, an $XX$ fault at the second swap location spreads to $X_5$ while also triggering the $Z$ syndrome.)  As $X_{5,7} \sim X_2 \widebar X$, the possible $X$ errors that can occur with a nontrivial $Z$ syndrome are all distinguishable, and similarly for the $Z$ errors.

\subsubsection{Three syndromes in parallel} \label{s:steaneparallel3syndromes}

\begin{figure}
\centering
\raisebox{0cm}{\includegraphics[scale=.769]{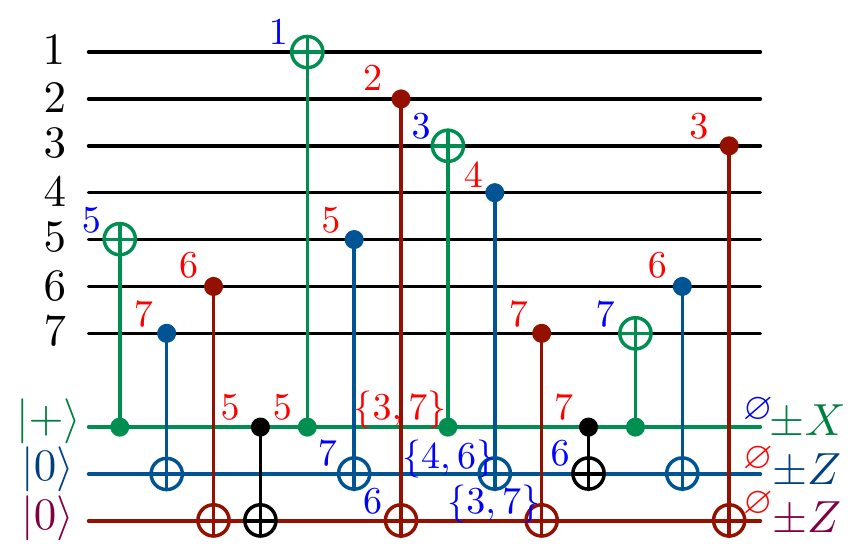}}
\caption{The same circuit as \figref{f:steaneparallel3syndromesunlabeled}, for measuring three Steane code syndromes in parallel, but with some $Z$ and $X$ failure locations marked in blue and red, respectively.} \label{f:steaneparallel3syndromes}
\end{figure}

In fact, with the circuit of \figref{f:steaneparallel3syndromes} three of the six syndromes can be extracted in parallel, flagging each other, using gates that are local according to the geometry of \figref{f:steane2d3ancillas}.  
Again, we have marked in blue $Z$ fault locations that can trigger the $X$ measurement.  Note that $Z_{4,6} \sim Z_2 \widebar Z$ and $Z_{3,7} \sim Z_4 \widebar Z$, so the possible $Z$ errors that can occur with a nontrivial $X$ syndrome (namely, $\identity, Z_1, Z_{4,6}, Z_3, Z_{3,7}, Z_5, Z_6, Z_7$) are all distinguishable.  
We have marked in red $X$ fault locations that can trigger one or both $Z$ measurements.  Note that both $X_4$ and $X_{3,7} \sim X_4 \widebar X$ are possible, and these errors are not distinguishable.  However, $X_4$ can only occur with $+,-$ $Z$ measurements, while $X_{3,7}$ can only occur with $-,-$ $Z$ measurements; so a special correction rule is only needed in the latter case.  

This circuit's increased parallelism is particularly useful when qubit initialization or measurement is slow.  
If CNOTs on different qubits can be applied in parallel, then the circuit uses one initialization round, six rounds of CNOTs, and one measurement round.  
It uses three extra qubits, fewer than the seven qubits needed for Steane's method for one-shot measurement of three syndromes.  
A symmetrical circuit works for the other three syndromes.  

\begin{figure}
\centering
\raisebox{0cm}{\includegraphics[scale=.476]{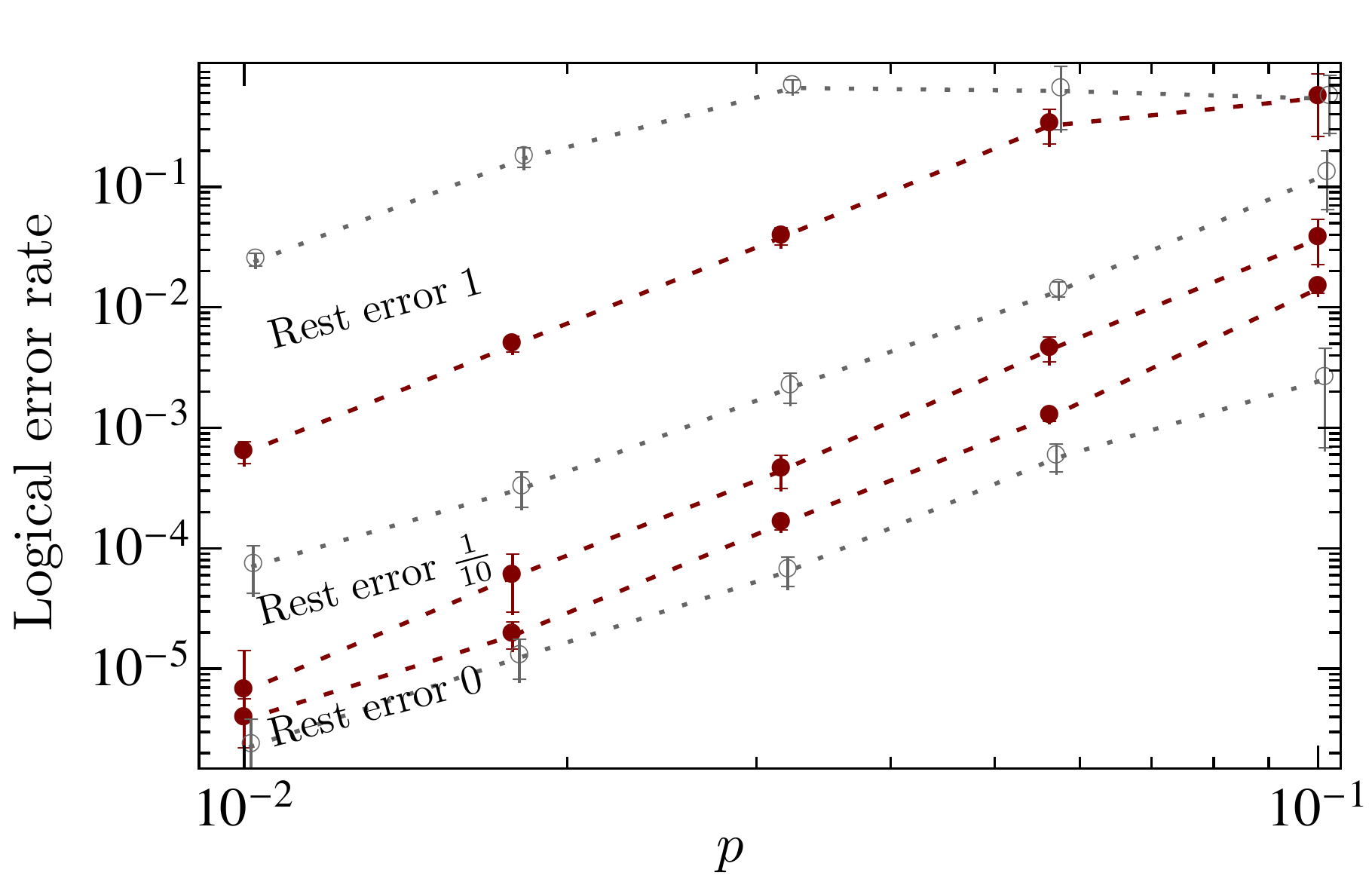}}
\caption{Logical error rates for simulated error detection; using the parallel syndrome-extraction circuit of \figref{f:steaneparallel3syndromes}, in red, and the sequential syndrome-extraction circuit of \figref{f:flagerrorcorrection}, in grey.
Errors are from a standard depolarizing noise model~\cite{Knill05}, with the error on resting qubits either $0$, $\tfrac{1}{10}$ or $1$ times the one-qubit marginal error rate of a CNOT gate.  
} \label{f:errordetectionplot}
\end{figure}

Observe that any single fault being correctable means that any two faults are detectable, so this circuit can also be used for an error-detection experiment.  Figure~\ref{f:errordetectionplot} plots the logical error rate for a simulated error-detection procedure, using either parallel or sequential syndrome extraction, for various relative values of the rest error~rate.  Note that since the procedure conditions on no detected errors, the overhead can be substantial.  

\subsection{Parallel syndrome extraction for the $\llbracket 4,2,2 \rrbracket$ code} \label{s:parallelsyndromeerasurecode}

We have tried to extend this method to other codes, i.e., to find circuits that extract two (or more) syndromes at once, using one syndrome to flag the other.  However, for most codes it is not easy.  

The $\llbracket 4,2,2 \rrbracket$ color code has a single plaquette on four vertices.  The (geometrically local) circuit below uses two extra qubits to extract both syndromes simultaneously, with one syndrome flagging the other: 
\begin{equation*}
\raisebox{0cm}{\includegraphics[scale=.769]{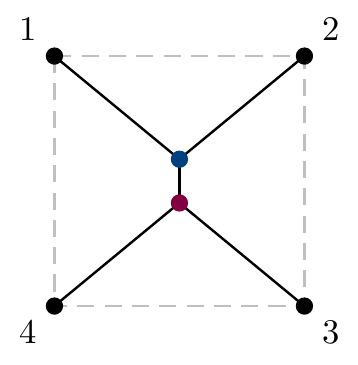}}
\qquad
\raisebox{0cm}{\includegraphics[scale=.769]{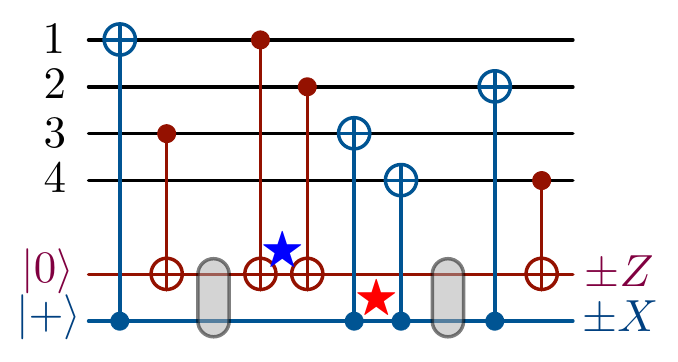}}
\end{equation*}
The circuit is fault tolerant in the sense that no single fault, including at the shaded swap locations, can cause an undetectable logical error on the encoded data.  (A $Z$ fault at $\color{blue}\bigstar$ propagates to $Z_{2,4}$, which is a logical operator---but it will be detected by the $X$ syndrome measurement.  Similarly for an $X$ fault at $\color{red}\bigstar$.)

\subsection{Parallel syndrome extraction with a shared flag} \label{s:sharedflag}

For the $\llbracket 5,1,3 \rrbracket$ code, stabilized by $XZZXI$ and its cyclic permutations, we can extract two syndromes at once using a third extra qubit as a flag for both: 
\begin{equation*}
\raisebox{0cm}{\includegraphics[scale=.769]{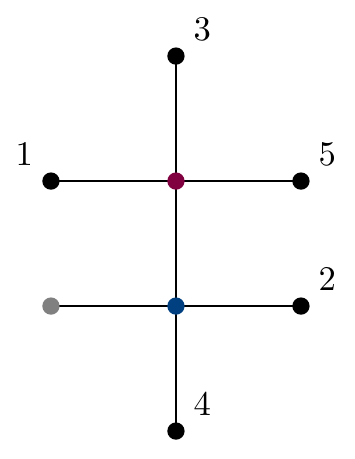}}
\,
\raisebox{0cm}{\includegraphics[scale=.769]{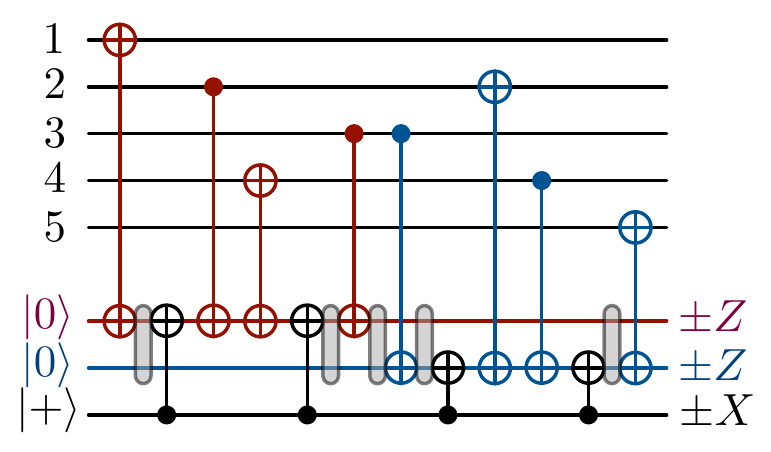}}
\end{equation*}
With this interaction geometry, the other two syndromes can also be extracted together.  We have not been able to find a fault-tolerant circuit that extracts two syndromes together without a third extra qubit.  

The $\llbracket 15,7,3 \rrbracket$ Hamming code is a self-dual CSS, perfect distance-three code.  (It can be seen as a three-dimensional color code, and is next in the family of $\llbracket 2^r - 1, 2^r - 1 - 2r, 3 \rrbracket$ Hamming codes.)  Its four $X$ and four $Z$ stabilizers are each given by the following parity-checks: 
\begin{equation*}
\begin{tabular}{c c c c c c c c c c c c c c c}
0&0&0&0&0&0&0&1&1&1&1&1&1&1&1\\
0&0&0&1&1&1&1&0&0&0&0&1&1&1&1\\
0&1&1&0&0&1&1&0&0&1&1&0&0&1&1\\
1&0&1&0&1&0&1&0&1&0&1&0&1&0&1
\end{tabular}
\end{equation*}
Index the qubits left to right from $1$ to $15$.  Observe that the columns are these numbers in binary.  

\begin{figure}
\centering
\begin{tabular}{c}
\subfigure[\label{f:1573parallel2syndromesZ}]{
\raisebox{0cm}{\includegraphics[scale=.769]{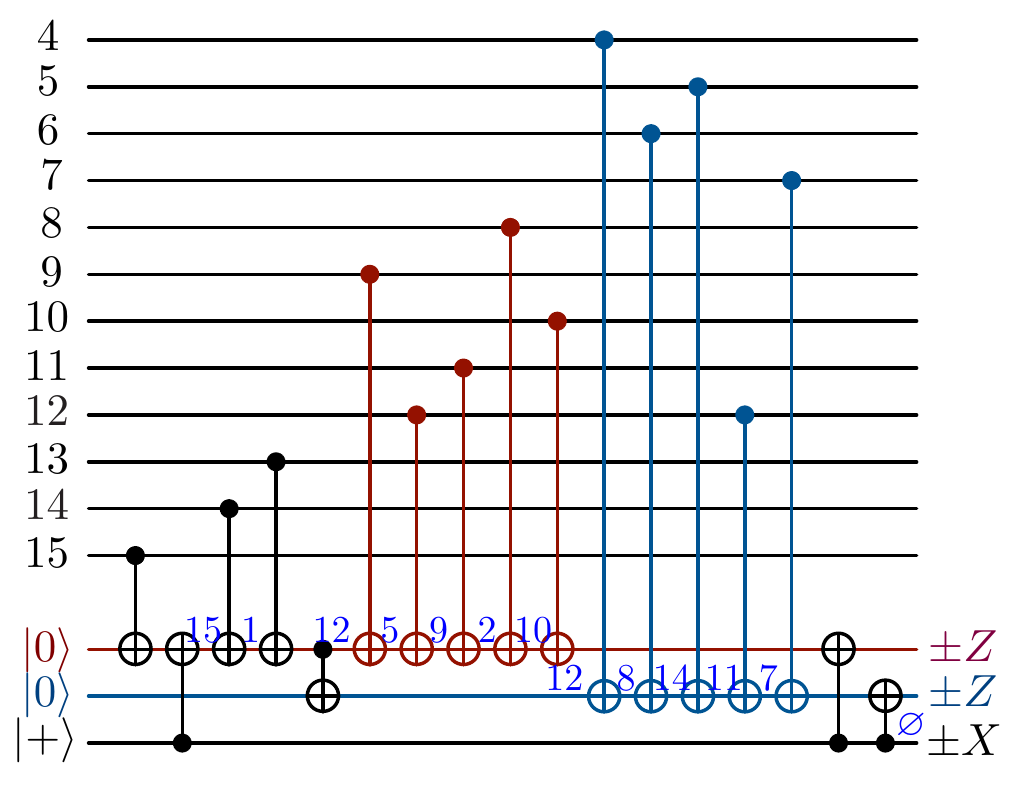}}
} \\
\subfigure[\label{f:1573parallel2syndromesdualXZ}]{
\raisebox{0cm}{\includegraphics[scale=.769]{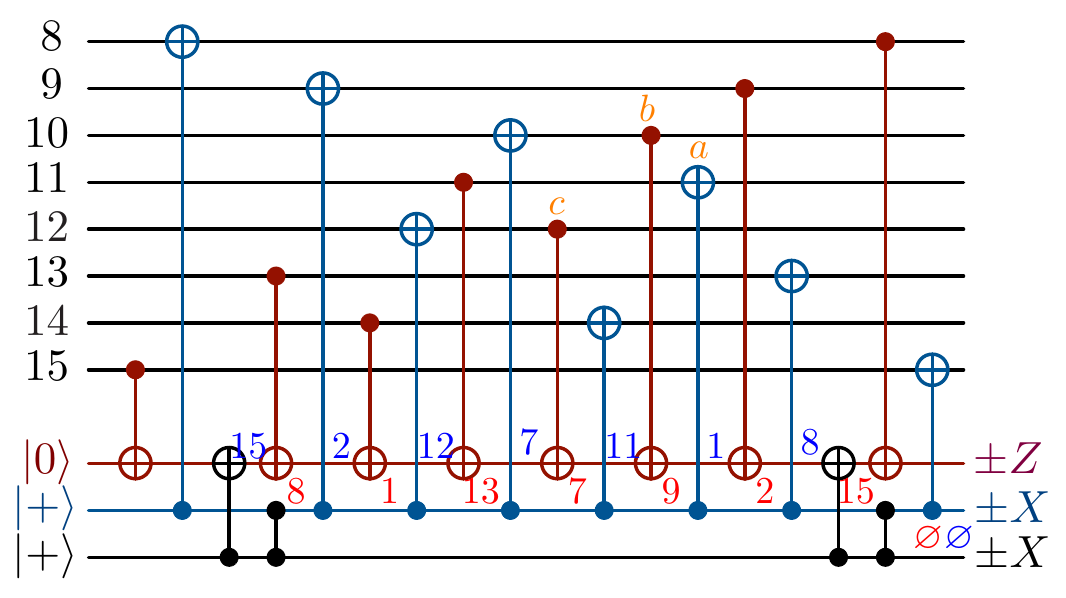}}
}
\end{tabular}
\caption{Circuits to extract in parallel two syndromes of the $\llbracket 15,7,3 \rrbracket$ Hamming code, with a shared flag.  
(a) Circuit to measure $Z_{8\ldots15}$ and $Z_{4\ldots7,12\ldots15}$, using fewer gates than sequential syndrome measurement.  
(b) Circuit to measure $Z_{8\ldots15}$ and $X_{8\ldots15}$.  
} \label{f:1573parallel2syndromes}
\end{figure}

The circuits in \figref{f:1573parallel2syndromes} each fault-tolerantly extract two syndromes at once, using a third extra qubit as a shared flag.  
The first circuit extracts two $Z$ syndromes, $Z_{8\ldots15}$ and $Z_{4\ldots7,12\ldots15}$, taking advantage of their overlap to reduce the required number of CNOT gates to $17$.  
$Z$ failure locations that trigger the flag are marked in blue with the syndrome of the error to which the failure propagates, e.g., a failure at either location marked ${\color{blue} 12}$ propagates to an error equivalent to $Z_{13,14,15} \sim Z_{12} \protect\widebar Z_S$, where $\protect\widebar Z_S$ is a logical $Z$ operator.  Observe that if the flag is triggered by a single fault, possible $Z$ errors are all distinguishable.  Single $Z$ faults that do not trigger the flag can cause at most a weight-one $Z$ error on the data.  

The circuit in \figref{f:1573parallel2syndromesdualXZ} extracts the syndromes of dual stabilizers, $Z_{8\ldots15}$ and $X_{8\ldots15}$.  It is more complicated to verify the fault tolerance of this circuit, which measures $Z_{8\ldots15}$ and $X_{8\ldots15}$.  For example, a $Z$ failure at ${\color{blue} 11}$ propagates to $Z_{8,9,10} \sim Z_{11} \protect\widebar Z_{S'}$ and triggers the flag.  However, a $ZX$ failure after CNOT $\color{Tangerine}a$ also triggers the flag, while propagating to $Z_{11} X_{13,15}$.  The $Z$ parts of the errors are inequivalent but indistinguishable, so the $X$ components must be used to distinguish these errors.  (The effects of a $YZ$ fault on CNOT $\color{Tangerine}b$ and an $XZ$ fault on CNOT $\color{Tangerine}c$ must also be distinguished.)

\section{Color code syndrome extraction without flags} \label{s:onequbitcolorcode}

For some surface codes, such as the $\llbracket 9,1,3 \rrbracket$ surface code, syndromes can be extracted fault tolerantly without flag qubits~\cite{TomitaSvore14surfacecode}.  One extra qubit is also enough for the $\llbracket 9,1,3 \rrbracket$ Bacon-Shor subsystem code~\cite{Bacon05operator, AliferisCross06BaconShorft}.  

\begin{figure}
\centering
\raisebox{-1.95cm}{\includegraphics[scale=.384]{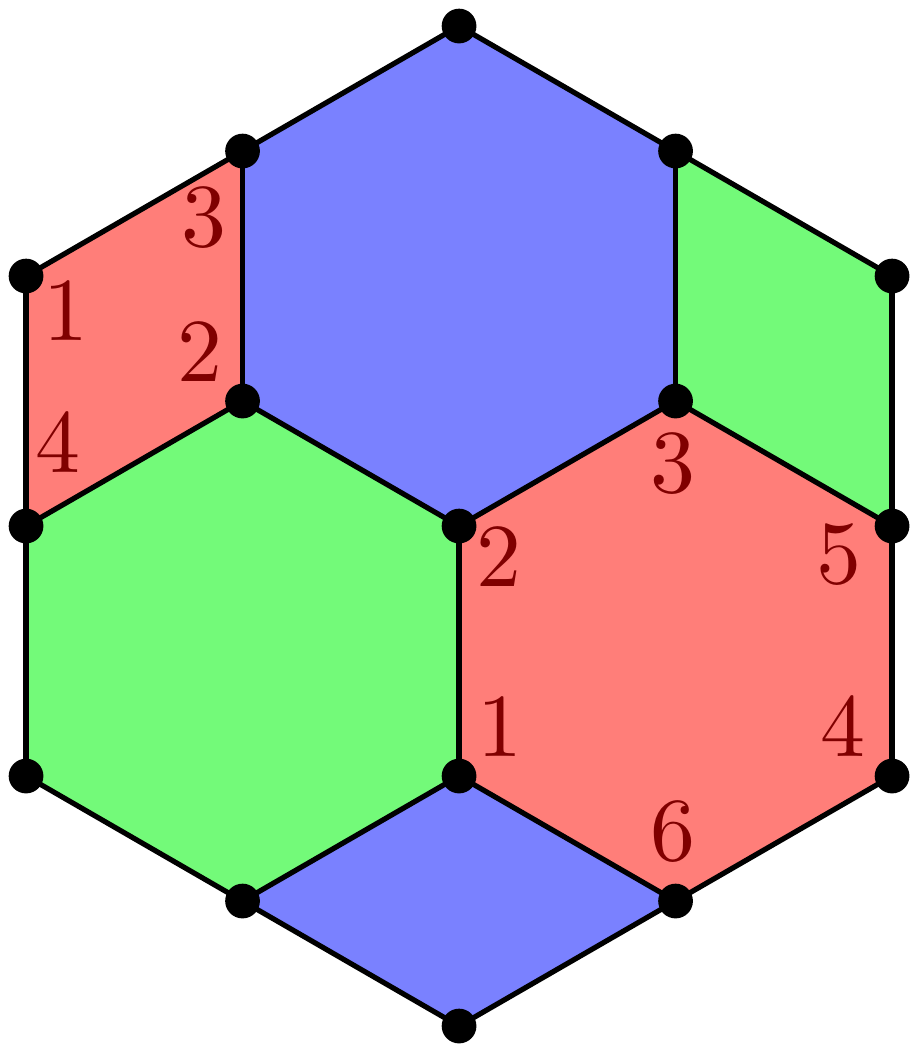}}
\caption{A $\llbracket 16,4,3 \rrbracket$ color code.} \label{f:1643colorcodesyndromeorder}
\end{figure}

Similarly, for a $\llbracket 16,4,3 \rrbracket$ color code, fault-tolerant error correction can be accomplished with one extra qubit per plaquette and naive, unflagged syndrome extraction.  
In \figref{f:1643colorcodesyndromeorder}, extract the syndromes for the two red plaquettes with CNOTs to or from the code qubits in the indicated order.  Other plaquette syndromes can be extracted with symmetrical orders.  A single $X$ fault during $X$ syndrome extraction can lead to a correlated $X$ error 
\begin{align*}
\raisebox{-.5cm}{\includegraphics[scale=.3]{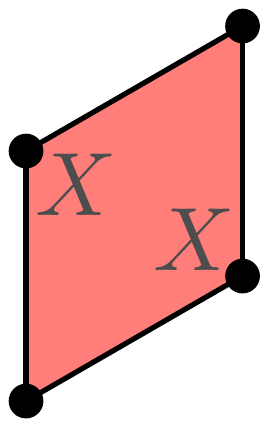}}
\;,\quad
\raisebox{-.5cm}{\includegraphics[scale=.3]{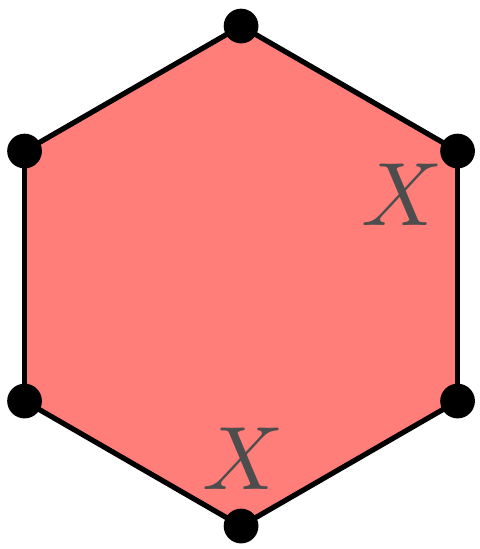}}
\;,\quad
\raisebox{-.5cm}{\includegraphics[scale=.3]{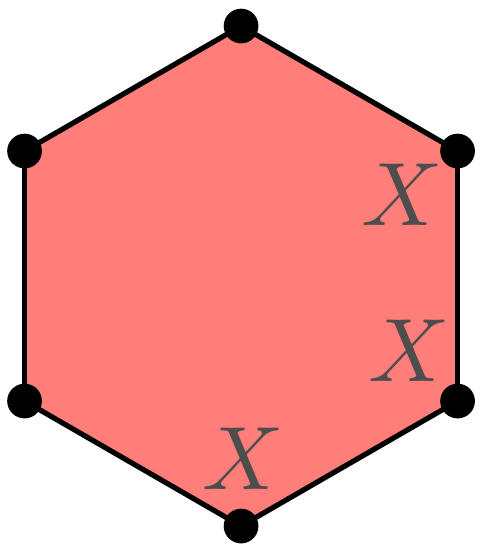}}
\;, \;\text{or} \quad
\raisebox{-.5cm}{\includegraphics[scale=.3]{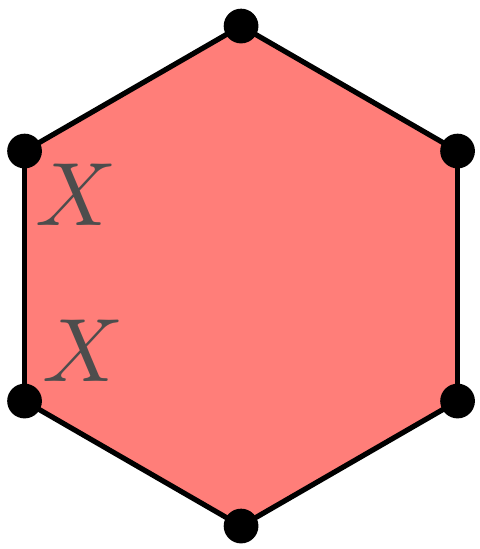}}
\end{align*}
These correlated errors, and the symmetrical errors on the other plaquettes, can all be detected and distinguished, from themselves and from weight-one errors.  (For example, the last triggers the two blue plaquette $Z$ stabilizers, which none of the other errors can do.)  Therefore they are all correctable.

\section{Merged color codes} \label{s:mergedcolorcodes}

\begin{figure}
\centering
\begin{align*}
4 \times \; \raisebox{-.4cm}{\includegraphics[scale=.192]{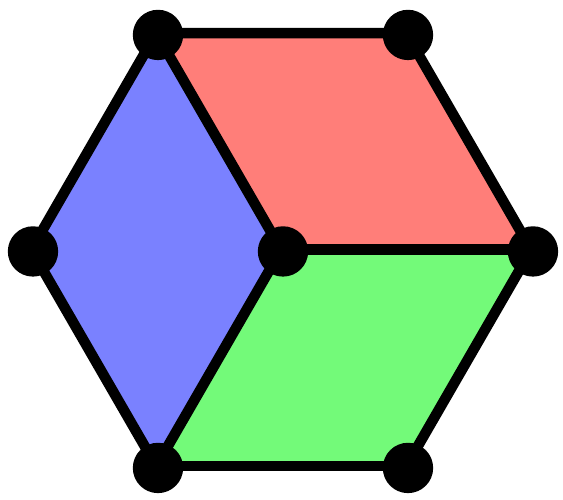}}
&\quad \overset{\raisebox{-.85cm}{\includegraphics[scale=.08]{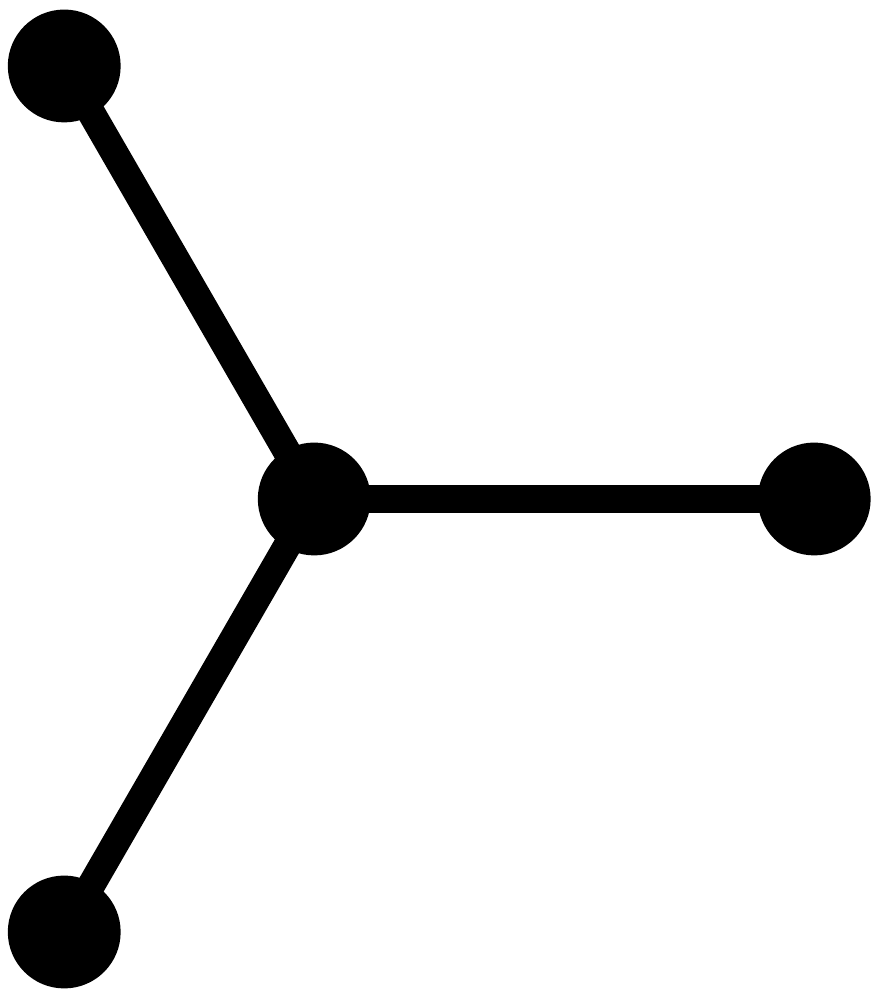}} }{\longrightarrow} \quad 
\raisebox{-1.2cm}{\includegraphics[scale=.192]{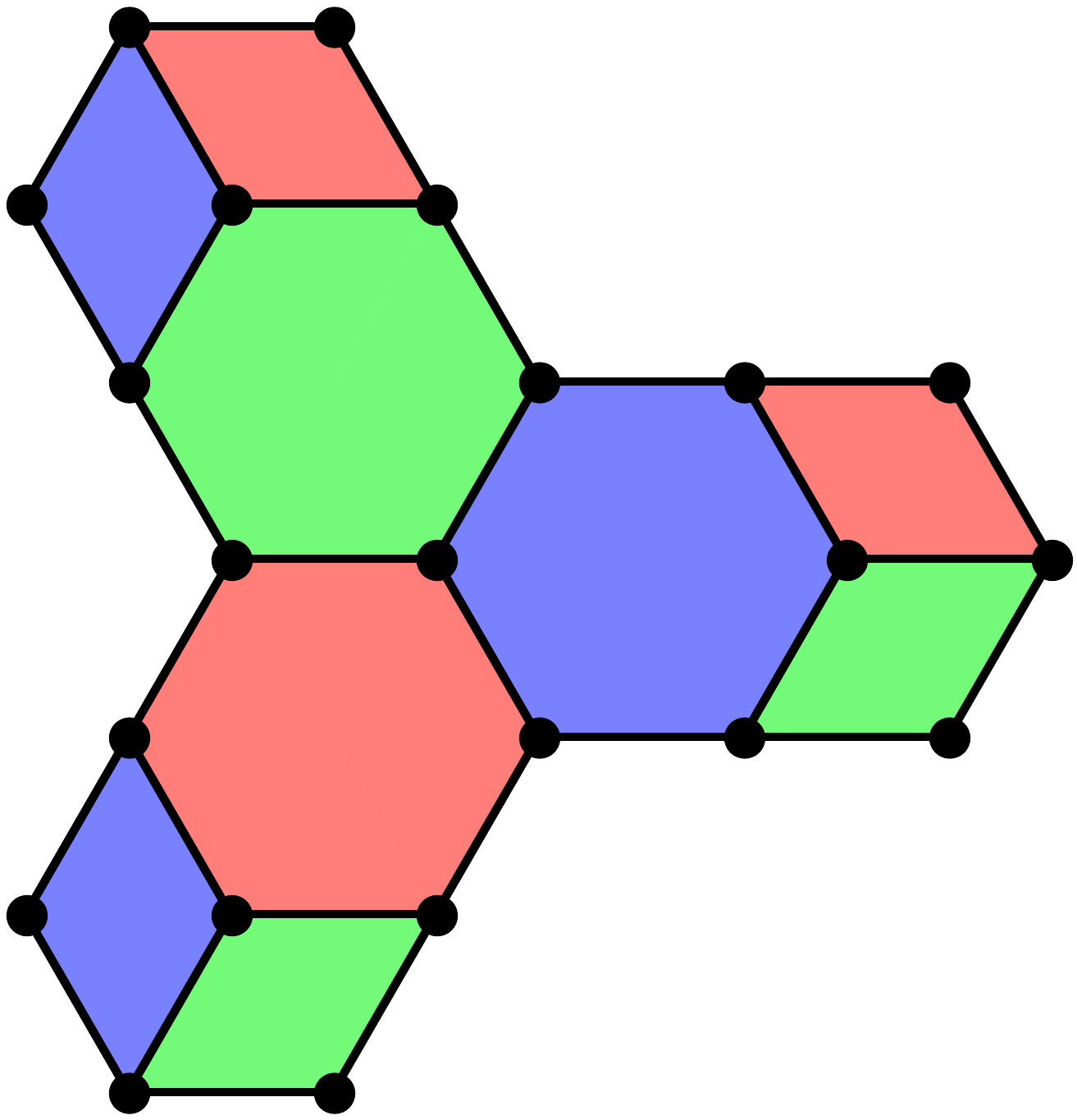}} \\
4 \times \; \raisebox{-.4cm}{\includegraphics[scale=.192]{images/steanecolor_small}}
&\quad \overset{\raisebox{-.85cm}{\includegraphics[scale=.08]{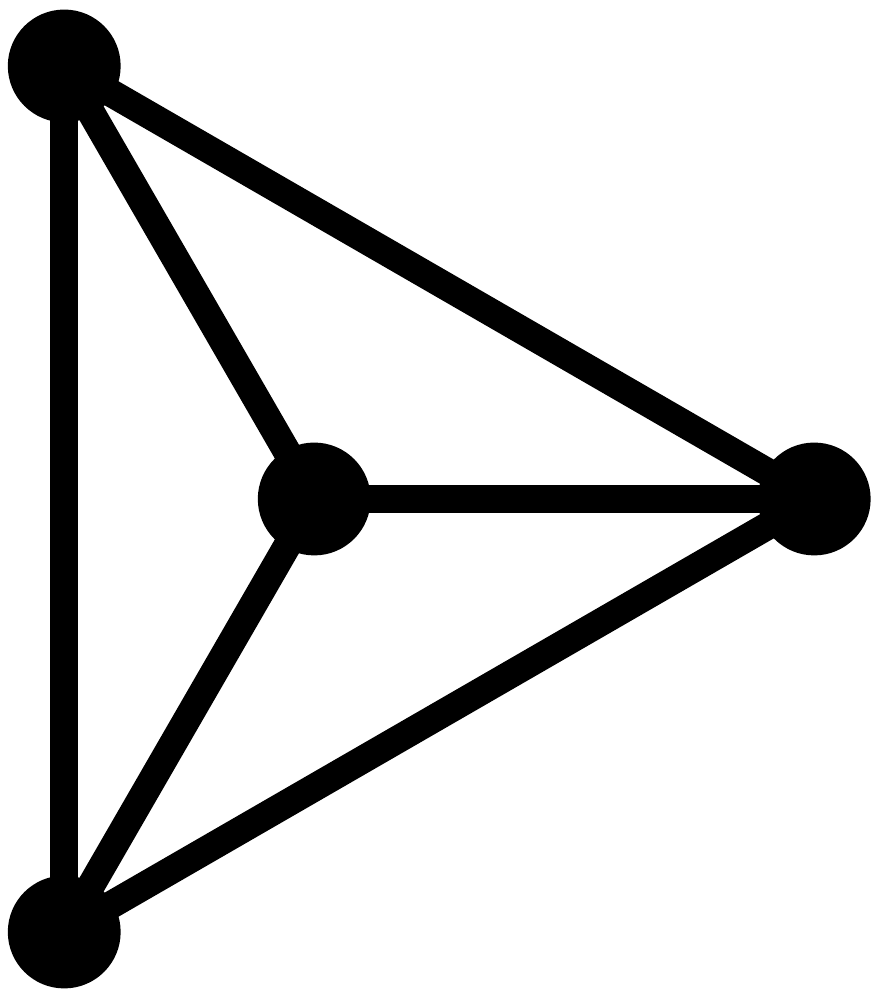}} }{\longrightarrow} \quad 
\hspace{0in}\raisebox{-1.2cm}{\includegraphics[scale=.192]{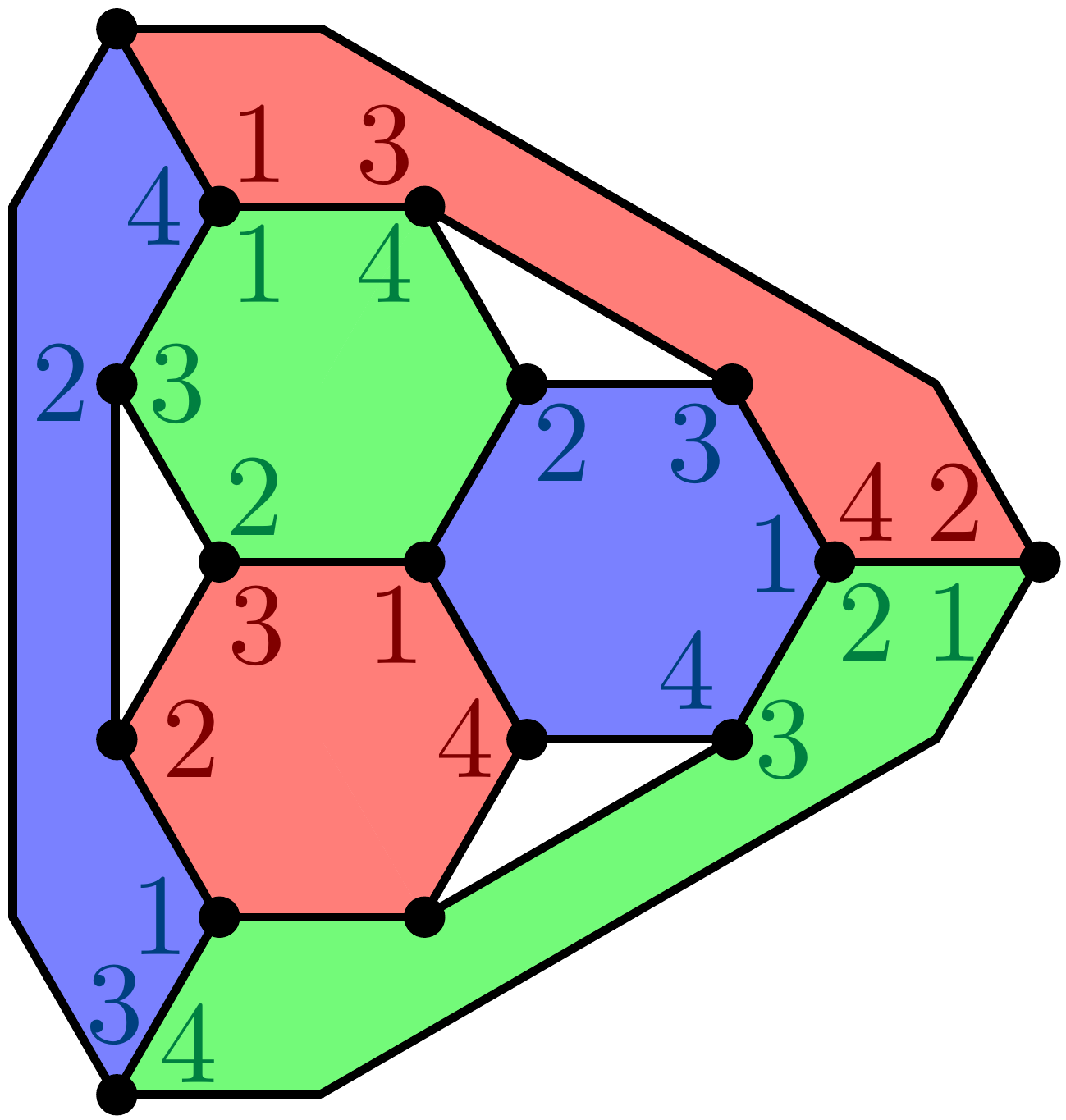}} 
\end{align*}
\caption{Four Steane code blocks can be merged to form $\llbracket 22,4,3 \rrbracket$ and $\llbracket 16,4,3 \rrbracket$ color codes.} \label{f:mergedcolorcodes}
\end{figure}

Color code blocks can be joined together to form larger codes.  For example, the $\llbracket 12,2,3 \rrbracket$ code described earlier arises from joining two Steane $\llbracket 7,1,3 \rrbracket$ code blocks (\figref{f:steanecodemerge1223}).  
Four Steane code blocks can be joined according to various graphs to get, for example, $\llbracket 22,4,3 \rrbracket$ and $\llbracket 16,4,3 \rrbracket$ color codes, shown in \figref{f:mergedcolorcodes}.  
As for the code of \figref{f:1643colorcodesyndromeorder}, the plaquette syndromes for this $\llbracket 16,4,3 \rrbracket$ code can be extracted into one extra qubit, without flags.  For each plaquette we have indicated a working CNOT gate order for extracting the syndrome---counting only to four because the last two qubits can go in either order.  

Higher-distance color codes can also be joined in this way; see \figref{f:fourdistance5codesmerge} below.

\smallskip
Let us study in more detail error correction for the $\llbracket 12,2,3 \rrbracket$ color code.  It provides more examples for the above syndrome-extraction techniques.  

\begin{figure}
\centering
\begin{tabular}{c@{$\quad$}c}
\subfigure[\label{f:1223colorcode}]{\raisebox{0cm}{\includegraphics[scale=.384]{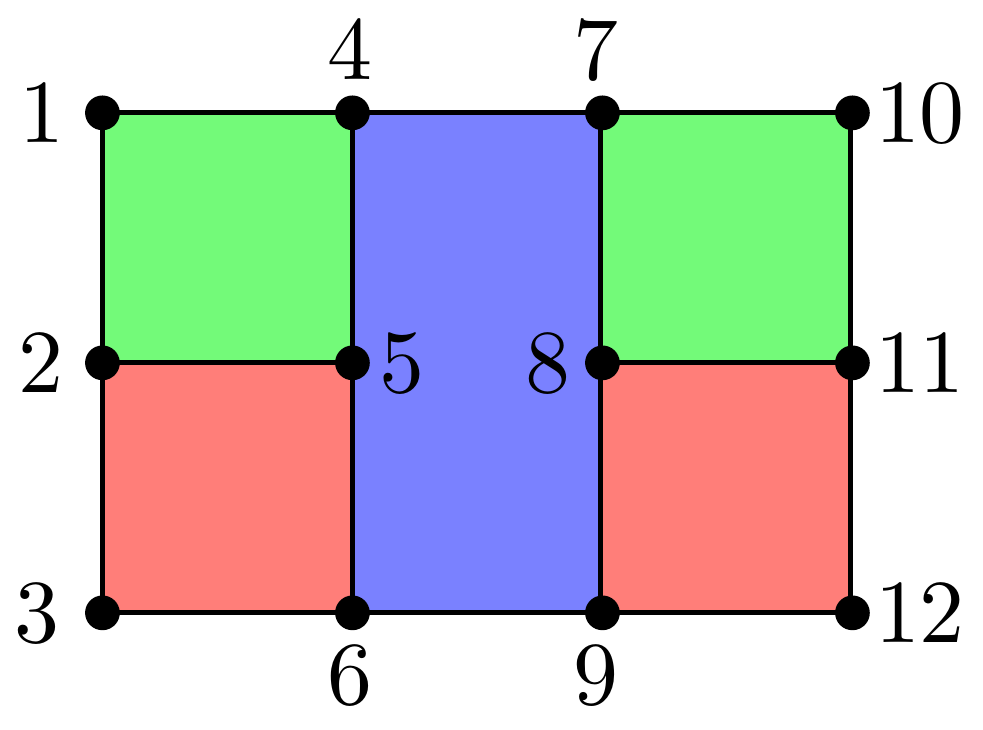}}} & 
\subfigure[\label{f:1223ancillas1}]{\raisebox{.36cm}{\includegraphics[scale=.384]{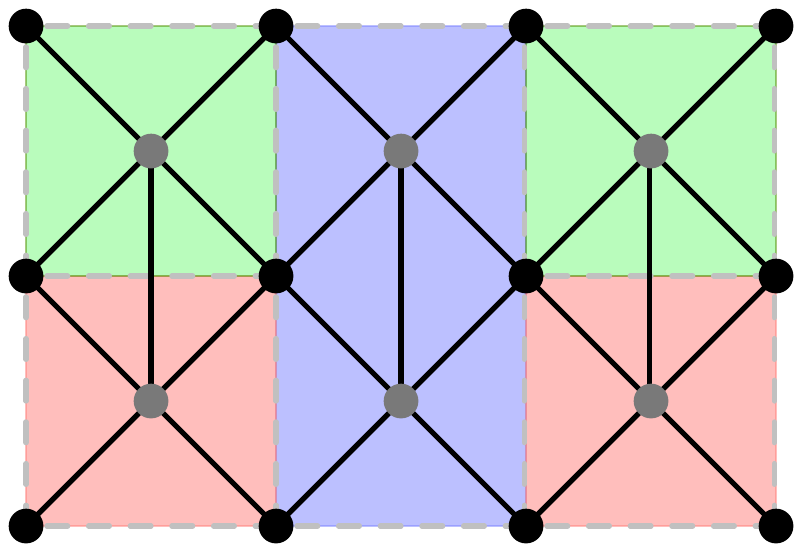}}} \\
\subfigure[\label{f:1223ancillas2}]{\raisebox{.36cm}{\hspace{-.15cm}\includegraphics[scale=.384]{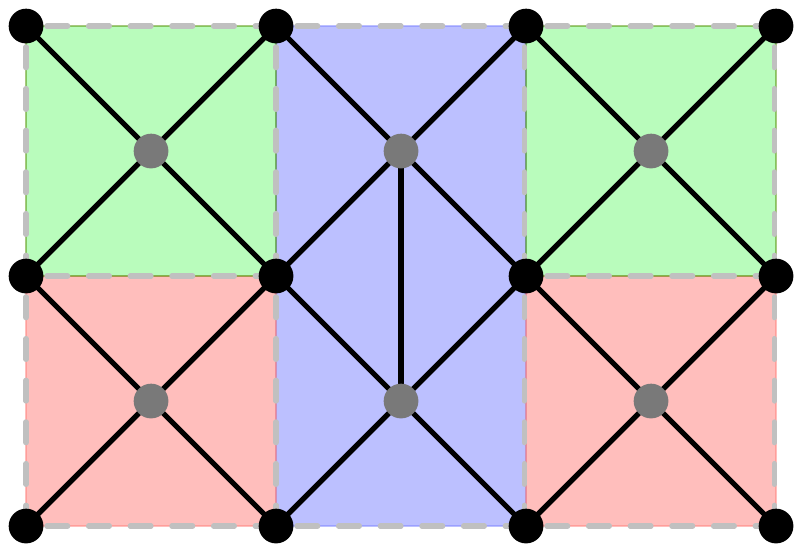}}} &
\subfigure[\label{f:1223ancillas3}]{\raisebox{.36cm}{\hspace{0cm}\includegraphics[scale=.384]{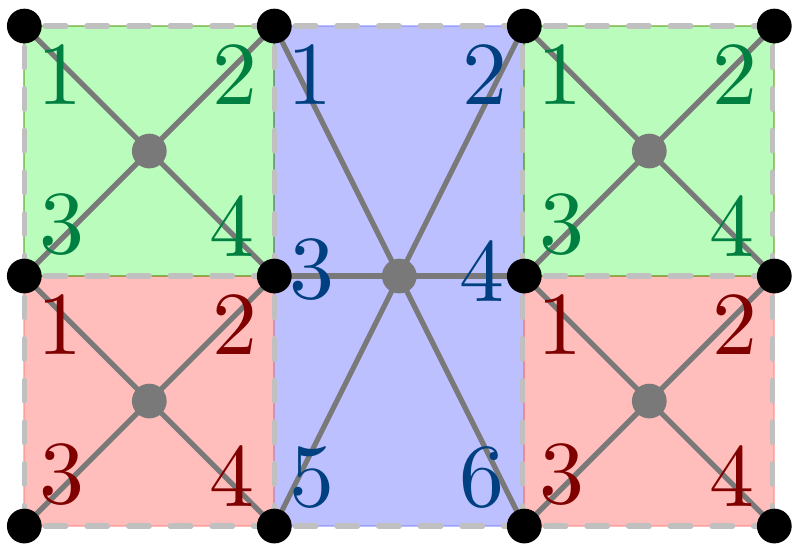}}}
\end{tabular}
\caption{The $\llbracket 12,2,3 \rrbracket$ code and three possible planar layouts for error correction.} \label{}
\end{figure}

Index the qubits as in \figref{f:1223colorcode}.  Logical operators can be chosen to be $\widebar X_1 = X_{1,2,3}, \widebar Z_1 = Z_{1,2,3}, \widebar X_2 = X_{10,11,12}, \widebar Z_2 = Z_{10,11,12}$.  

\begin{figure}
\centering
\begin{tabular}{c}
\subfigure[]{
\raisebox{0cm}{\includegraphics[scale=.769]{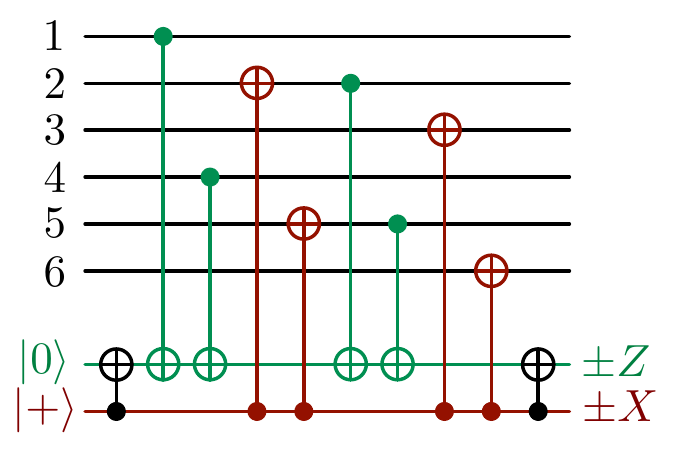}}
} \\
\subfigure[]{
\raisebox{0cm}{\includegraphics[scale=.769]{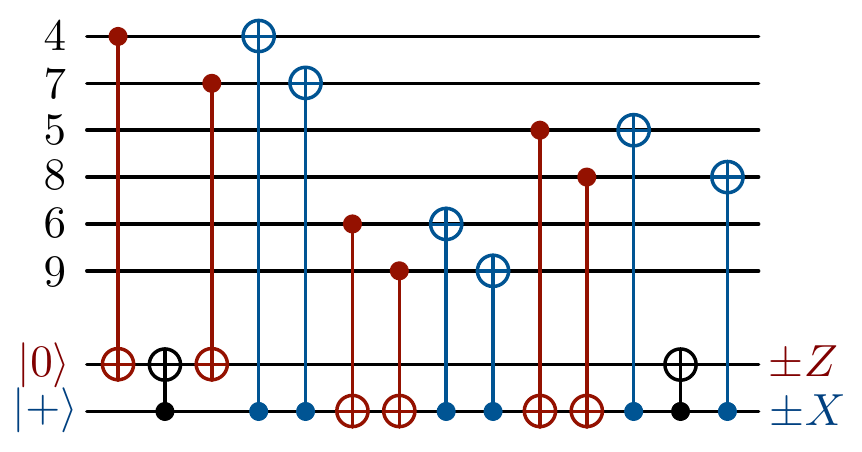}}
}
\end{tabular}
\caption{Circuits to extract $\llbracket 12,2,3 \rrbracket$ code syndrome pairs.  
} \label{f:1223parallel2syndromes}
\end{figure}

First consider the planar layout of \figref{f:1223ancillas1}, using six extra qubits.  
Using the circuits in \figref{f:1223parallel2syndromes}, the weight-four stabilizers can be measured in parallel, flagging each other, 
and the central weight-six stabilizers can be measured in parallel using the two enclosed qubits.  

In fact, fault-tolerant error correction is also possible with the sparser qubit connectivity pattern of \figref{f:1223ancillas2}.  This connectivity requires that the weight-four stabilizers be measured without flags, which takes some care.  There are essentially three inequivalent orders for the CNOT gates to measure a square plaquette, vertically, horizontally or diagonally: 
\begin{equation*}
\raisebox{0cm}{\includegraphics[scale=.244]{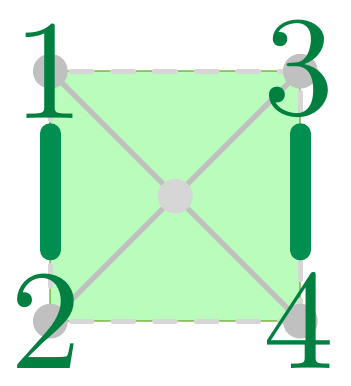}}
\qquad\quad
\raisebox{0cm}{\includegraphics[scale=.244]{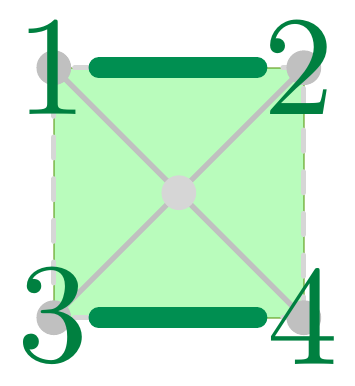}}
\qquad\quad
\raisebox{0cm}{\includegraphics[scale=.244]{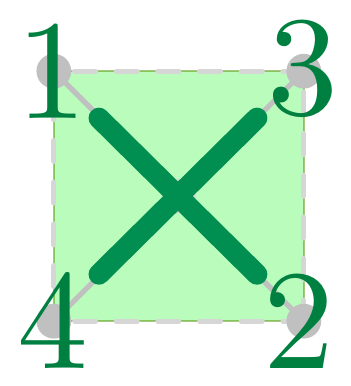}}
\end{equation*}
Applied to measure $Z_{1,2,4,5}$ in \figref{f:1223colorcode}, for example, these orders respectively lead to the correlated errors $Z_{1,2}$, $Z_{1,4}$ and $Z_{1,5}$.  As $Z_{1,2} \sim Z_3 \widebar Z$ and $Z_{1,5} \sim Z_6 \widebar Z$, the vertical and diagonal correlated errors are uncorrectable.  This leaves the horizontal order.  However, the horizontal correlated errors from the left and right square plaquettes are indistinguishable, e.g., $Z_{1,4} \sim Z_{7,10} \widebar Z_{1,2}$.  Therefore the left plaquettes' $Z$ stabilizers cannot be measured together with those of the right plaquettes.  

Instead, one can measure together $Z_{1,2,4,5}, Z_{2,3,5,6}$, $X_{7,8,10,11}, X_{8,9,11,12}$, then measure together $X_{4\ldots9}$ and $Z_{4\ldots9}$ (flagging each other, as above), then measure together $X_{1,2,4,5}, X_{2,3,5,6}$, $Z_{7,8,10,11}, Z_{8,9,11,12}$, etc.: 
\begin{equation*}
\raisebox{-.4cm}{\includegraphics[scale=.192]{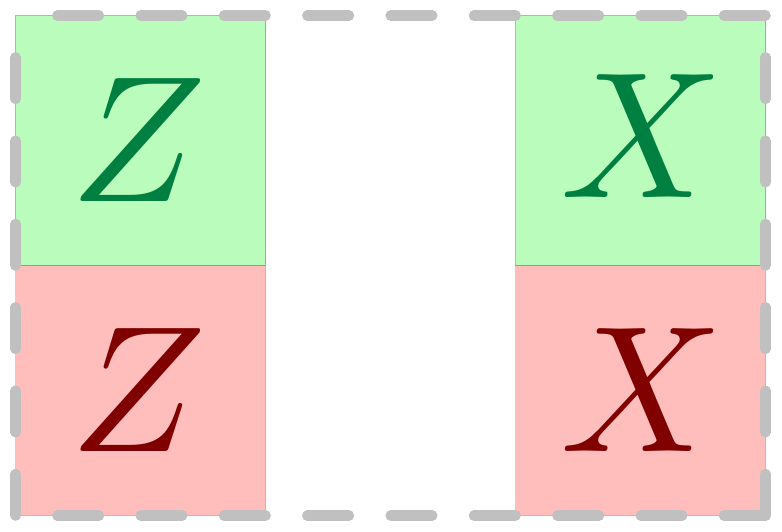}} \,,\;
\raisebox{-.4cm}{\includegraphics[scale=.192]{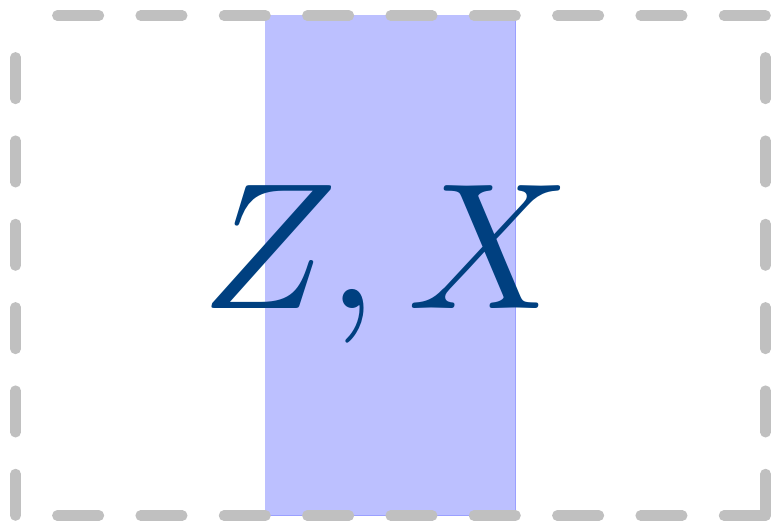}} \,,\;
\raisebox{-.4cm}{\includegraphics[scale=.192]{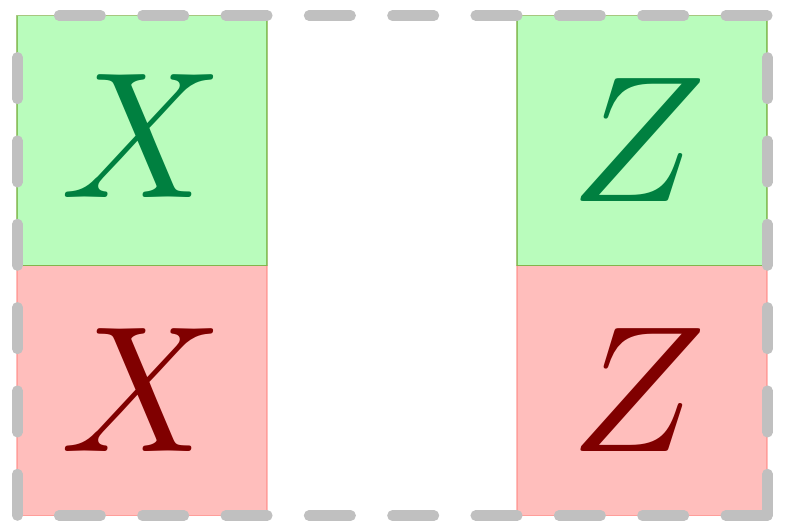}} \,,\;
\raisebox{-.4cm}{\includegraphics[scale=.192]{images/1223altmeasurement2}}
\end{equation*}
Correlated errors created in the first round, like $Z_{1,4}$ and $X_{7,10}$, are caught and distinguished by the weight-6 stabilizer measurements.  
(If $Z_{4\ldots9}$ is triggered, then possible $X$ errors are $\identity, X_4, \ldots, X_9, X_{5,8}, X_{5,8,9}, X_{5,6,8}, X_{5,6,8,9} \sim X_{4,7}$ and $X_{7,10}$, and these are all distinguishable.  Similarly, if $X_{4\ldots9}$ is triggered, then possible $Z$ errors are $\identity, Z_4, \ldots, Z_9, Z_{5,8}, Z_{5,8,9}, Z_{5,6,8}, Z_{5,6,8,9} \sim Z_{4,7}$ and $Z_{1,4}$, and these are all distinguishable.)  

Finally, \figref{f:1223ancillas3} shows a qubit layout in which the central weight-six stabilizers $X_{4\ldots9}$ and $Z_{4\ldots9}$ must also be extracted without flags.  For simplicity, extract all the syndromes on the left, then center, then right, etc.: 
\begin{equation*}
\raisebox{-.4cm}{\includegraphics[scale=.192]{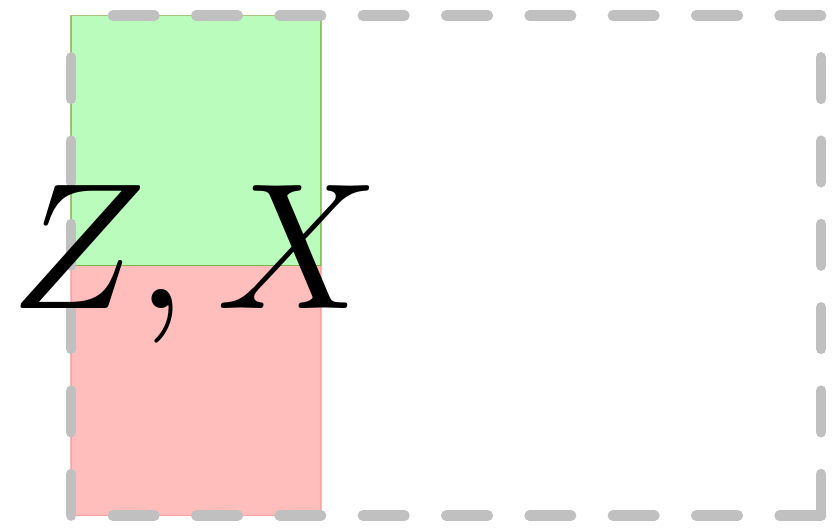}} \,,\;
\raisebox{-.4cm}{\includegraphics[scale=.192]{images/1223altmeasurement2}} \,,\;
\raisebox{-.4cm}{\includegraphics[scale=.192]{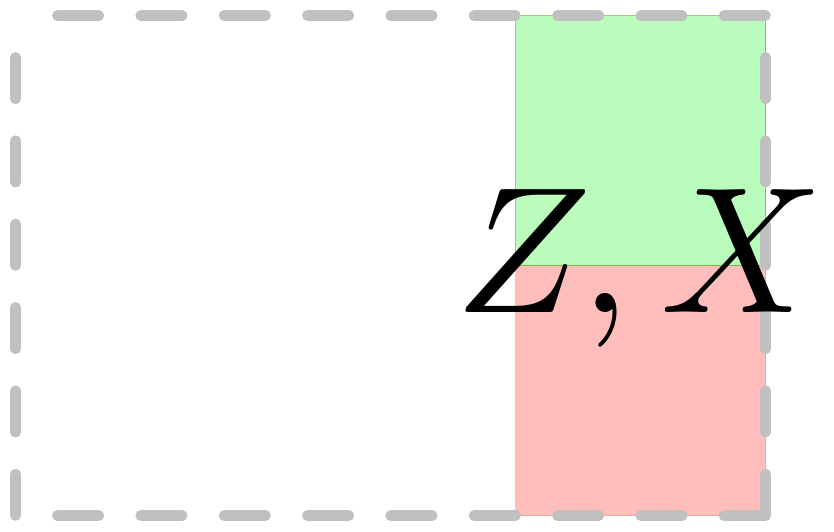}} \,,\;
\raisebox{-.4cm}{\includegraphics[scale=.192]{images/1223altmeasurement2}}
\end{equation*}
Apply the CNOT gates for each plaquette in the order given in \figref{f:1223ancillas3}.  Before starting the right syndrome measurements, possible $X$ errors are $\identity, X_1, \ldots, X_{12}$ and the correlated errors $X_{4,7}, X_{4,5,7}, X_{6,9}$ (from measuring $X_{4\ldots9}$).  These errors are all distinguishable, and the correlated errors will be detected by the right syndrome measurements $Z_{7,8,10,11}, Z_{8,9,11,12}$.  Similarly for $Z$ errors.  Before starting the center syndrome measurements again, possible $X$ errors are $\identity, X_1, \ldots, X_{12}$ and $X_{7,10}$.  Once more, these errors are distinguishable, and $X_{7,10}$ will be detected by $Z_{4\ldots9}$.  The key point in this scheme is that it is not necessary to distinguish $X_{1,4}$ from $X_{7,10}$, because the center syndrome measurement $Z_{4\ldots9}$ goes between left and right $X$ syndrome measurements.

\section{Steane code error correction with no extra qubits} \label{s:steanesteaneerrorcorrection}

Inspired by the construction of the $\llbracket 12,2,3 \rrbracket$ color code from two Steane $\llbracket 7,1,3 \rrbracket$ codewords, and by the parallel syndrome-extraction procedures studied above, we have designed an error-correction method that uses no extra qubits, provided that there are at least two Steane code blocks.  Roughly, combining the code blocks frees up two qubits into which two syndromes can be measured.  

Consider the following circuit $\cal C$ on two code blocks: 
\begin{align*}
\raisebox{0cm}{\includegraphics[scale=.769]{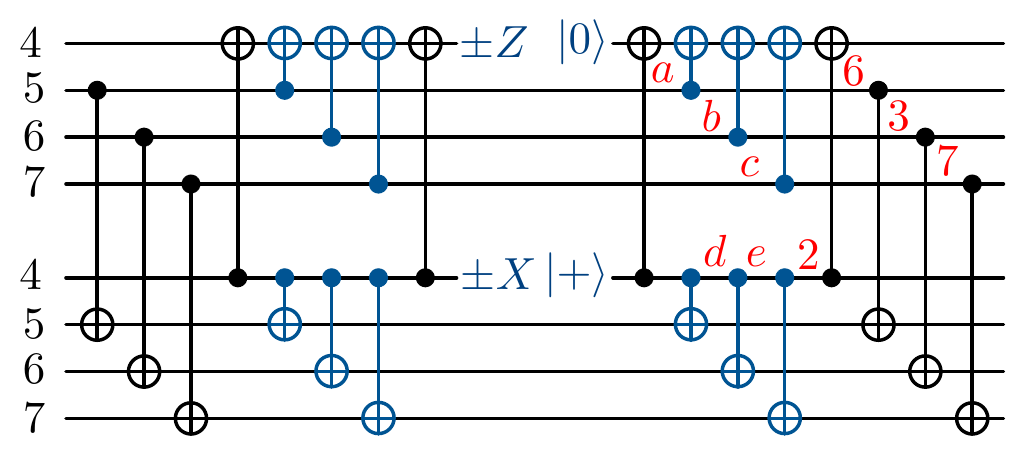}} 
\end{align*}
(Marked locations will be used below.)  
The structure of the gates is indicated below.  
\begin{align*}
\raisebox{0cm}{\includegraphics[scale=.384]{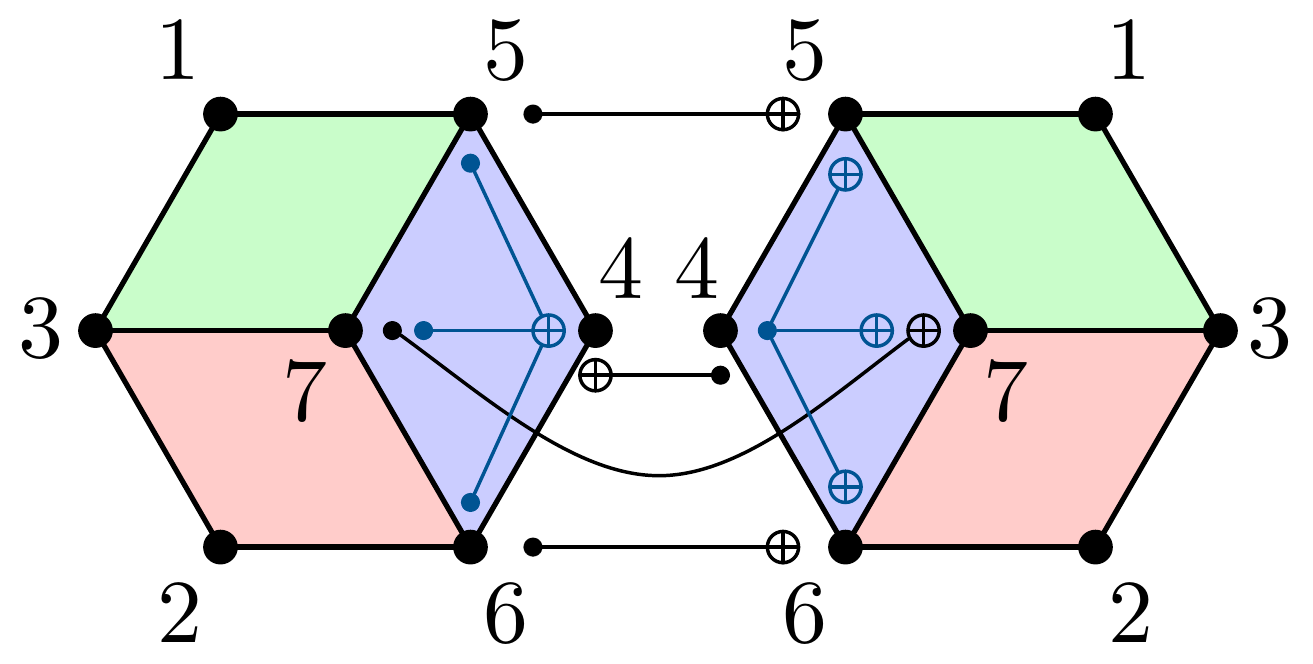}} 
\end{align*}
The blue CNOT gates alone extract the syndromes for $Z_{4,5,6,7} \otimes \identity$ and $\identity \otimes X_{4,5,6,7}$, then reinitialize those stabilizers.  The black gates formally commute with the blue gates and cancel out; they are present in order to catch faults during syndrome extraction.  

{ \noindent \hrulefill \\  
\centering \textbf{Error-correction procedure} \\ } \smallskip
\noindent
Repeat six times: 
\begin{enumerate}[leftmargin=*]
\item 
Use $\cal C$ to measure $Z_{4,5,6,7} \otimes \identity$ and $\identity \otimes X_{4,5,6,7}$.  
\item 
Switch the code blocks and rotate each by one notch, so the blue plaquette moves into the red plaquette, which moves into the green, into the blue.  Thus the qubits are permuted as $1 \mapsto 4 \mapsto 2 \mapsto 1$, $3 \mapsto 5 \mapsto 6 \mapsto 3$, $7 \mapsto 7$.  (This qubit reindexing is just meant to simplify our presentation, so we assume it is perfect.)  
\end{enumerate}
Without errors, the six rounds give all twelve syndromes: 
\begin{align*}
1. \; Z_{4,5,6,7} \otimes \identity ,\, \identity \otimes X_{4,5,6,7} \;\;\;\; 4. \; X_{4,5,6,7} \otimes \identity ,\, \identity \otimes Z_{4,5,6,7} \\
2. \; X_{1,3,5,7} \otimes \identity ,\, \identity \otimes Z_{1,3,5,7} \;\;\;\; 5. \; Z_{1,3,5,7} \otimes \identity ,\, \identity \otimes X_{1,3,5,7} \\
3. \; Z_{2,3,6,7} \otimes \identity ,\, \identity \otimes X_{2,3,6,7} \;\;\;\; 6. \; X_{2,3,6,7} \otimes \identity ,\, \identity \otimes Z_{2,3,6,7}
\end{align*}

If any measurement is nontrivial, then start over and use the syndromes to diagnose and correct the error.  

\vspace{-.5\baselineskip}
\noindent
\hrulefill
\medskip

A single fault in the syndrome-extraction circuit $\cal C$ can lead to a correlated error in the output.  For example, a $Y$ fault at $d$ spreads undetected to $-i X_4 \otimes (Z_4 X_5)$.  Worse, an $X$ fault at location $a$ spreads to $X_{4,5} \otimes X_5$.  In the Steane code, the two-qubit error $X_{4,5} \sim X_1 \widebar X$ is indistinguishable from $X_1$.  A naive error-correction procedure, that treated the two code blocks separately, would therefore introduce a logical $X$ error.  Once the two code blocks have interacted via~$\cal C$, it is important that the error-correction rules take into account possible correlations in errors across the blocks.  
Note that it is not possible for a single gate fault in $\cal C$ to create an error that is trivial in one code block and is uncorrectable (has an $X$ or $Z$ component of weight at least two) in the other block.  

Let $\S_X$ be the set of errors 
\begin{equation*}
\begin{gathered}
\identity \otimes \identity \\
X_j \otimes \identity, \; \identity \otimes X_j \\
E_j = X_j \otimes X_j \\
\begin{aligned}
E_a &= X_6 \otimes X_{2,6} & E_d &= X_6 \otimes X_2
\\ E_b &= X_3 \otimes X_{2,3} & E_e &= X_{2,7} \otimes X_2 
\\ E_c &= X_7 \otimes X_{1,4} & E_0 &= X_{1,5} \otimes X_5 
\end{aligned}
\end{gathered}
\end{equation*}
where $j \in \{1, \ldots, 7\}$.  Let $\S_Z$ be the same as $\S_X$ except with the two code blocks swapped and $X$ replaced by~$Z$.  

The following claim asserts that the error-correction procedure is fault tolerant~\cite{AliferisGottesmanPreskill05}, using $\S_X$ and $\S_Z$ as induction invariants for the input and output errors.  

\begin{claim} \label{t:errorcorrection}
The error-correction procedure satisfies: 
\begin{enumerate}[leftmargin=*]
\item 
If there are no gate faults in error correction, then the outputs lie in the code space.  
Moreover, if the $X$ and $Z$ components of the input error lie in $\S_X$ and $\S_Z$, respectively, and there are no faults in error correction, then the error is corrected.  
\item 
If the inputs are in the codespace and there is at most one fault, then the output error's $X$ and $Z$ components lie in $\S_X$ and~$\S_Z$.  
\end{enumerate}
\end{claim}

Note that $\S_X$ does not include arbitrary pairs of one-qubit errors $X_i \otimes X_j$.  Even though such an error has weight one on both code blocks, it is not necessarily correctable together with the other errors in $\S_X$; for example, $X_{2,6} = X_4 \widebar X$, so $E_a$ and $X_6 \otimes X_4$ are indistinguishable, inequivalent errors.  This is acceptable because two one-qubit errors is in general a second-order event.  

\begin{proof}[Proof of \claimref{t:errorcorrection}]
The first statement follows because the errors in $\S_X$ all have different $Z$ syndromes.  This implies that once an error in $\S_X$ is detected, a faultless syndrome-extraction procedure is enough to correct it.  $Z$ errors are symmetrical.  

Second, assume that the inputs are in the codespace and there is at most one fault in the six steps of error correction.  We assert that the output error's $X$ component lies in~$\S_X$.  (Then by symmetry the error's $Z$ component lies in $\S_Z$.)  

If a single $X$ fault occurs in~$\cal C$, then whether or not the $Z$ measurement detects it, it can only propagate to an error in $\S_X$.  In particular, faults at the locations marked $2, 3, 6, 7, a, \ldots, e$ propagate to $E_2, \ldots, E_e$, respectively, after the qubit permutation.  Thus the assertion holds provided the fault occurs in the last of the six syndrome-extraction rounds.  

Now consider if the input to a syndrome-extraction round is in~$\S_X$, from a previous fault.  Of the correlated errors, only $E_1, E_2, E_3$ and~$E_b$ are undetected by the $Z$ measurement.  After the qubit permutation, these become, respectively, $E_4, E_1, E_5$ and $E_0$.  Therefore the output still lies in~$\S_X$, as desired.  
\end{proof}

Is this construction useful?  Probably not in a $14$-qubit experiment.  With $14$ qubits, one could encode two logical qubits into the $\llbracket 12,2,3 \rrbracket$ code and use the two remaining qubits for error correction---or have two $\llbracket 5,1,3 \rrbracket$ codewords with two error-correction qubits for each codeword.  The construction becomes useful when there are a larger number of logical qubits.  One can envision laying out $\llbracket 7,1,3 \rrbracket$ codewords in a 2D lattice.  Storing each logical qubit in its own code block makes applying fault-tolerant operations more convenient.  To correct errors, pair each codeword up with a neighbor.

\section{Error-correcting other codes without extra qubits} \label{s:codecodeerrorcorrection}

The method for error-correcting two Steane codewords without extra qubits extends to some other CSS codes.  (It does not seem to work for the non-CSS $\llbracket 5,1,3 \rrbracket$ code, essentially because in the analogous circuit to~$\cal C$ a $ZX$ fault after a CNOT gate is not caught and is uncorrectable.)

\subsection{$\llbracket n,n-2,2 \rrbracket$ erasure code}

For an even integer $n$, the $\llbracket n,n-2,2 \rrbracket$ erasure code is a color code with a single plaquette; the stabilizers are $X^{\otimes n}$ and $Z^{\otimes n}$.  A circuit on two code blocks extracts the syndromes of $Z^{\otimes n} \otimes \identity$ and $\identity \otimes X^{\otimes n}$ without extra qubits: 
\begin{equation*}
\raisebox{0cm}{\includegraphics[scale=.769]{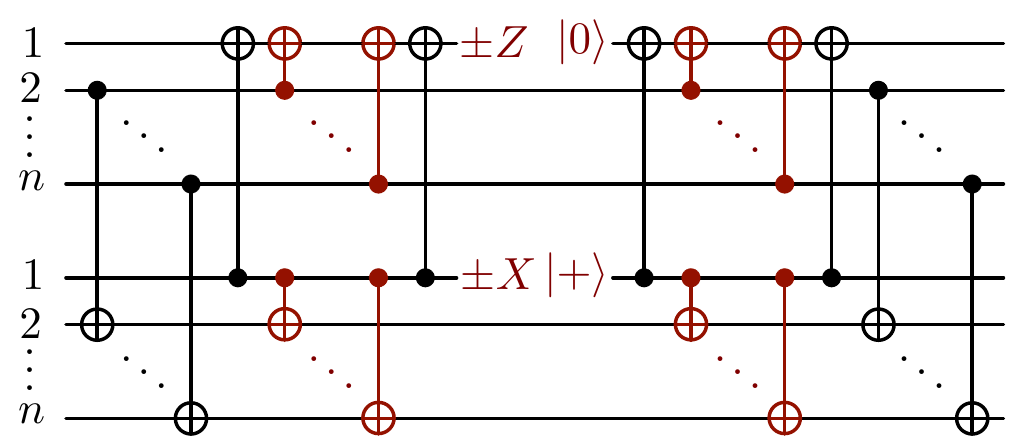}} 
\end{equation*}
It is fault tolerant in the sense that no one fault can lead to an undetectable logical error.  For example, a single $X$ fault cannot propagate to an error with even weight on both blocks.  
Note though that in order to measure two syndromes, the circuit uses $6 n - 2$ CNOTs, substantially more than the $2 (n+2)$ CNOTs used by the two-extra-qubit flagged error-correction procedure of~\cite{ChaoReichardt17errorcorrection}.

\subsection{$\llbracket 15,7,3 \rrbracket$ Hamming code} 

\begin{figure}
\centering
\raisebox{0cm}{\includegraphics[scale=.769]{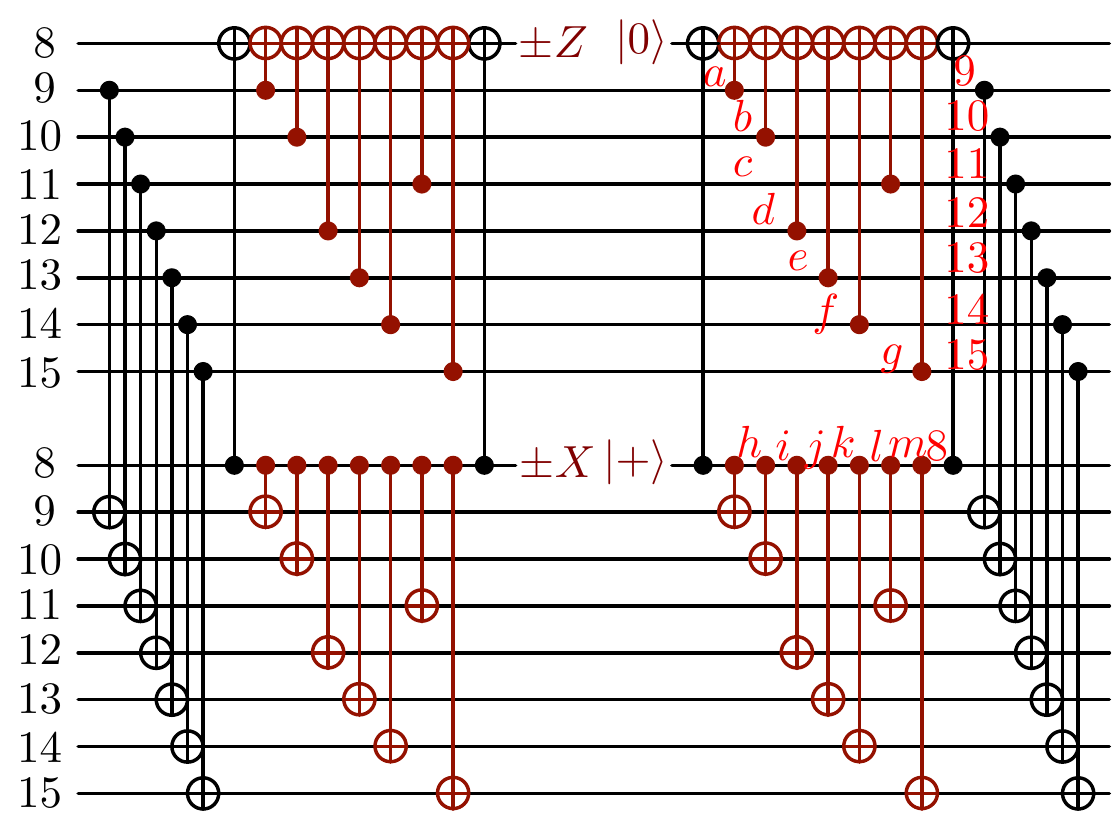}} 
\caption{Circuit for measuring the syndromes of $Z_{8,\ldots,15} \otimes \identity$ and $\identity \otimes X_{8,\ldots,15}$ on two $\llbracket 15,7,3 \rrbracket$ Hamming code blocks.} \label{f:hamminghammingerrorcorrectioncircuit}
\end{figure}

Let $\cal C$ be the circuit on two code blocks given in \figref{f:hamminghammingerrorcorrectioncircuit}.  It uses $46$ CNOT gates.  
The red CNOT gates alone extract the syndromes for $Z_{8,\ldots,15} \otimes \identity$ and $\identity \otimes X_{8,\ldots,15}$, then reinitialize those stabilizers.  The black gates formally commute with the other gates and cancel.  

{ \noindent \hrulefill \\ 
\centering \textbf{Error-correction procedure} \\ } \smallskip
\noindent
Repeat eight times: 
\begin{enumerate}[leftmargin=*]
\item 
Use $\cal C$ to measure $Z_{8,\ldots,15} \otimes \identity$ and $\identity \otimes X_{8,\ldots,15}$.  
\item 
Switch the code blocks and permute the qubits by $\sigma = (1,8,4,2)(3,9,12,6)(5,10)(7,11,13,14)(15)$.  
\end{enumerate}
The eight rounds give all $16$ code syndromes.  If any syndrome measurement is nontrivial, then measure all syndromes and use them to diagnose and correct the error.  

\vspace{-.5\baselineskip}
\noindent
\hrulefill
\medskip

Let $\S_X$ be the set of errors displayed in \figref{f:hamminghamminginductionerrors} mapped forward by the permutation~$\sigma$.  Let $\S_Z$ be the same as $\S_X$ except with the two code blocks swapped and $X$ replaced by~$Z$.  
Just as for the $\llbracket 7,1,3 \rrbracket$ Steane code, $\S_X$ and $\S_Z$ are induction invariants for the input and output errors.  The following observations imply \claimref{t:errorcorrection} for this error-correction procedure: 
\begin{itemize}[leftmargin=*]
\item 
Any $X$ fault in $\cal C$ propagates to an error in $\S_X$.  
\item 
The errors in $\S_X$ all have different $Z$ syndromes.  (Hence once an error in $\S_X$ is detected, a faultless syndrome-extraction procedure is enough to correct it.)  
\item 
Any error in $\S_X$ that is not detected by $Z_{8,\ldots,15} \otimes \identity$ (which is $\sigma(\identity \otimes Z_{1,3,\ldots,15})$) is mapped by $\sigma$ to another error in $\S_X$.  (For example, error ${\color{red}b}$ is not detected by $\identity \otimes Z_{1,3,\ldots,15}$, and is mapped to the error~$b'$.  Error~$b'$ is detected by $\identity \otimes Z_{1,3,\ldots,15}$.)  
\end{itemize}

\begin{figure}
\centering
\def\red{red}
\begin{equation*}
\begin{gathered}
(\identity, \identity), \qquad (X_\alpha, \identity), \qquad (\identity, X_\alpha) \\
\begin{aligned}
{\color{\red}\alpha} &:  X_\alpha \otimes X_\alpha
\\ {\color{\red}a} &: (X_{8,9} \sim X_1 \widebar X, X_9)
\\ {\color{\red}b} &: (X_{8,10} \sim X_2 \widebar X, X_{10})
& b' &: (X_5, X_1 \widebar X) 
\\ {\color{\red}c} &: (X_{8,11} \sim X_3 \widebar X, X_{11})
\\ {\color{\red}d} &: (X_{8,12} \sim X_4 \widebar X, X_{12})
& d' &: (X_6, X_2 \widebar X) 
\\ && d'' &: (X_1 \widebar X, X_3) 
\\ {\color{\red}e} &: (X_{8,13} \sim X_5 \widebar X, X_{13})
\\ {\color{\red}f} &: (X_{8,14} \sim X_6 \widebar X, X_{14})
& f' &: (X_7, X_3 \widebar X) 
\\ {\color{\red}g} &: (X_{8,15} \sim X_7 \widebar X, X_{15})
\\ {\color{\red}h} &: (X_8, X_9)
\\ {\color{\red}i} &: (X_8, X_{11\ldots15} \sim X_3 \widebar X)
\\ {\color{\red}j} &: (X_8, X_{13,14,11,15} \sim X_{15} \widebar X)
\\ {\color{\red}k} &: (X_8, X_{8,11,14,15} \sim X_2 \widebar X)
& k' &: (X_1 \widebar X, X_4) 
\\ && k'' &: (X_2, X_8 \widebar X) 
\\ && k''' &: (X_4 \widebar X, X_1) 
\\ {\color{\red}l} &: (X_8, X_{8,11,15} \sim X_{12} \widebar X)
& l' &: (X_6 \widebar X, X_4) 
\\ && l'' &: (X_2, X_3 \widebar X) 
\\ {\color{\red}m} &: (X_8, X_{8,15} \sim X_7 \widebar X)
\end{aligned}
\end{gathered}
\end{equation*}
\caption{$\S_X$, the induction invariant for $X$ errors, consists of these errors mapped forward by~$\sigma$.  Here $\alpha \in \{1, \ldots, 15\}$.  The red labels ${\color{\red}\alpha}$ and ${\color{\red}a}, \ldots, {\color{\red}m}$ indicate locations in \figref{f:hamminghammingerrorcorrectioncircuit} at which an $X$ fault can cause the error.} \label{f:hamminghamminginductionerrors}
\end{figure}

\section{Steane code error correction with one extra qubit}

Another example of correcting two code blocks at once will help illustrate the concept.  The following circuit uses one extra qubit to measure the parity of the syndromes for two code blocks, $Z_{4,5,6,7} \otimes Z_{4,5,6,7}$: 
\begin{equation*}
\raisebox{0cm}{\includegraphics[scale=.769]{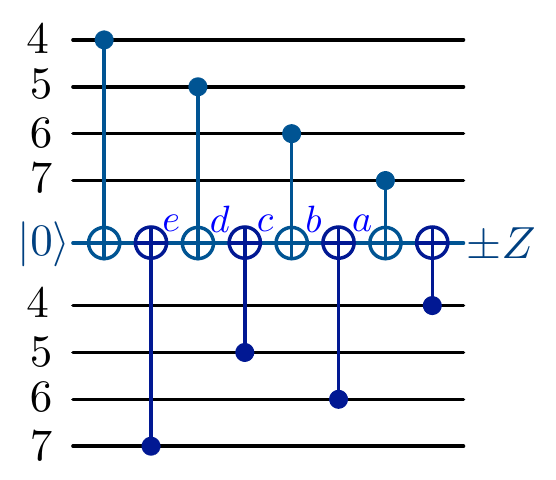}} 
\end{equation*}
With one $Z$ fault, possible $Z$ errors on the two blocks are 
\begin{equation*}
\def\blue{blue}
\begin{gathered}
(\identity, \identity), \qquad (Z_\alpha, \identity), \qquad (\identity, Z_\alpha) \\
\begin{aligned}
{\color{\blue}a} &: (Z_7, Z_4)
& {\color{\blue}d} &: (Z_{6,7}, Z_7)
\\ {\color{\blue}b} &: (Z_7, Z_{4,6} \sim Z_2 \widebar Z)
& {\color{\blue}e} &: (Z_4, Z_7)
\\ {\color{\blue}c} &: (Z_{6,7} \sim Z_1 \widebar Z, Z_{4,6})
\end{aligned}
\end{gathered}
\end{equation*}
where $\alpha \in \{1, \ldots, 7\}$ and the error labels $\color{blue}a$ through~$\color{blue}e$ correspond to the circuit's marked failure locations.  Note that for each error $(Z_S, Z_T)$, the product $Z_S Z_T$ has nontrivial Steane code syndrome.  The errors are thus all detectable by the syndrome parities $X_{4,5,6,7} \otimes X_{4,5,6,7}, X_{2,3,6,7} \otimes X_{2,3,6,7}$ and $X_{1,3,5,7} \otimes X_{1,3,5,7}$, and are distinguishable by the separate syndromes $X_{4,5,6,7} \otimes \identity, \identity \otimes X_{4,5,6,7}, \ldots$.  Therefore, although these errors can have weight two or more, they are correctable by a procedure that first measures the parities of corresponding pairs of syndromes and, on detecting an error measures the two code's syndromes separately without flags.  

Essentially, one code block is used to flag errors in the other.  This method uses fewer CNOTs than the scheme in \secref{s:steanesteaneerrorcorrection}, eight versus $22$.  However, it similarly increases the size of an error-correction rectangle compared to correcting the code blocks separately, and it cannot correct arbitrary weight-one errors in both code blocks, e.g., $X_\alpha \otimes X_\alpha$ is not detected.  (The error-correction procedure hence fails the AGP fault-tolerance conditions $0$ and~$0'$~\cite{AliferisGottesmanPreskill05}.  This is okay for fault-tolerant computation with one encoding level, but is not enough for analyzing two or more concatenation levels.)

\section{Higher-distance codes}

\begin{figure}
\centering
\begin{equation*}
\raisebox{-.18cm}{\includegraphics[scale=.192]{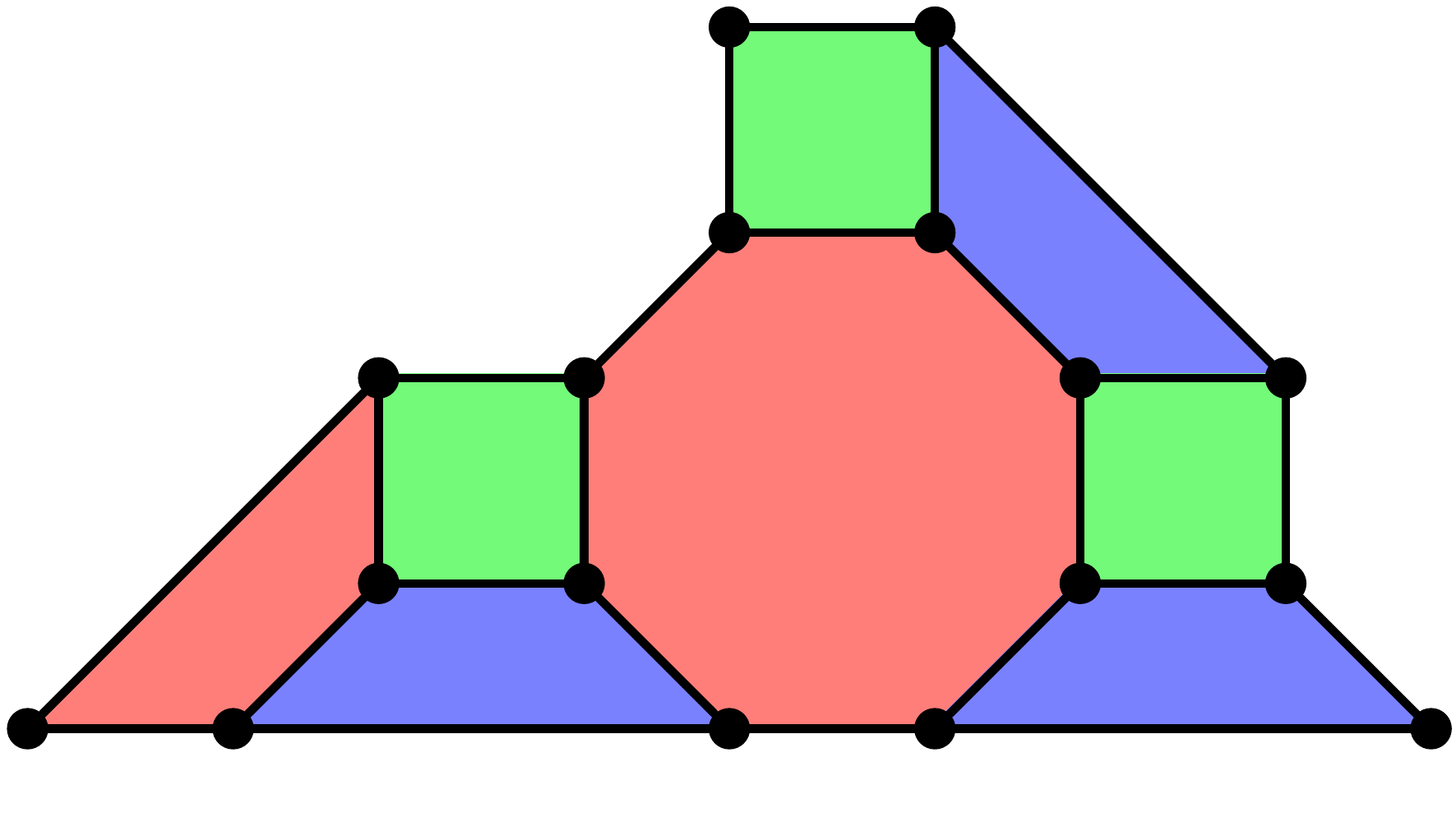}} 
\qquad
\raisebox{0cm}{\includegraphics[scale=.192]{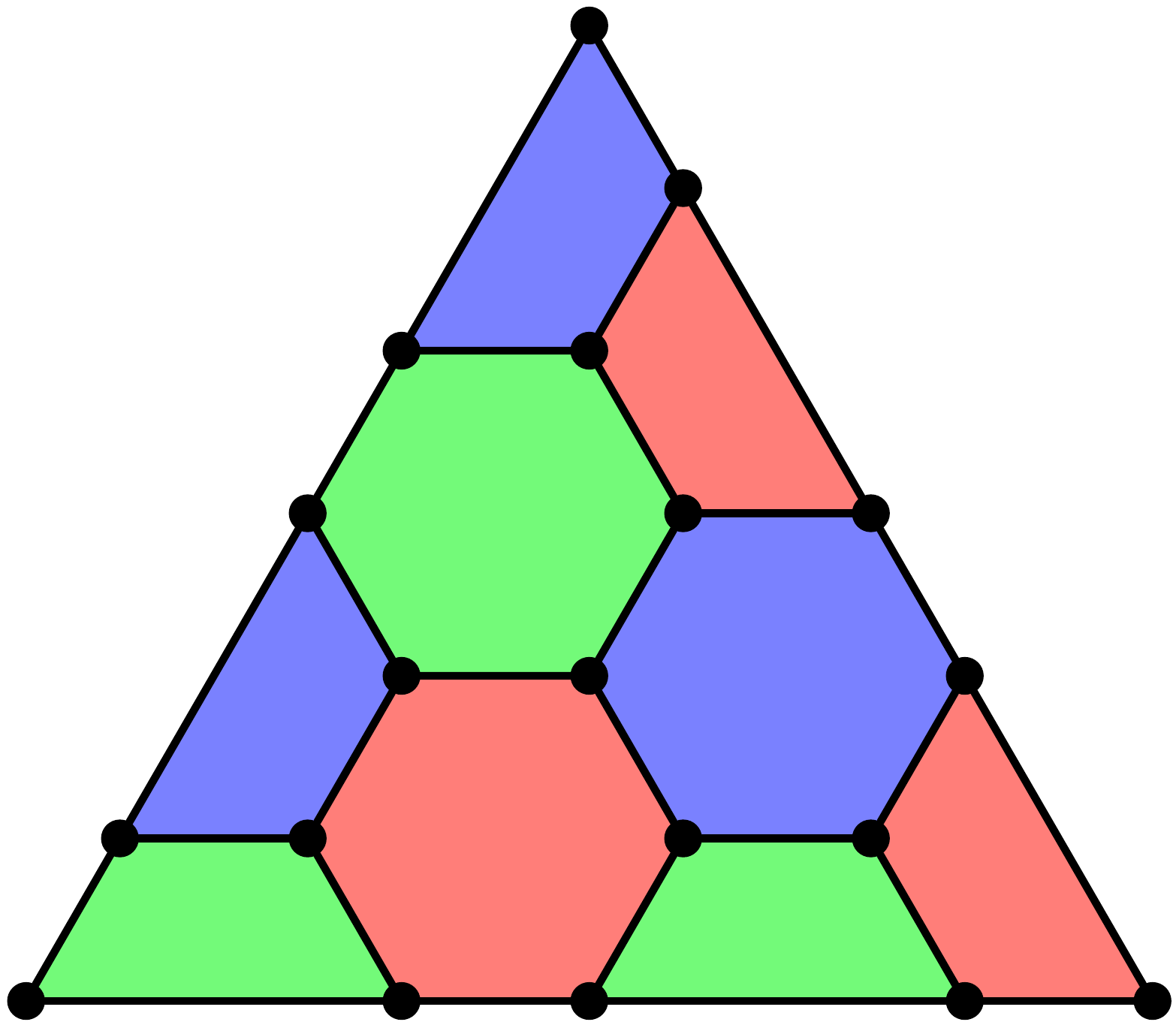}} 
\end{equation*}
\caption{$\llbracket 17,1,5 \rrbracket$ and $\llbracket 19,1,5 \rrbracket$ color codes.  Flagged syndrome extraction for these codes was studied in~\cite{ChamberlandBeverland17flags}.  
} \label{f:1715colorcode}
\end{figure}

\begin{figure}
\centering
\raisebox{0cm}{\includegraphics[scale=.384]{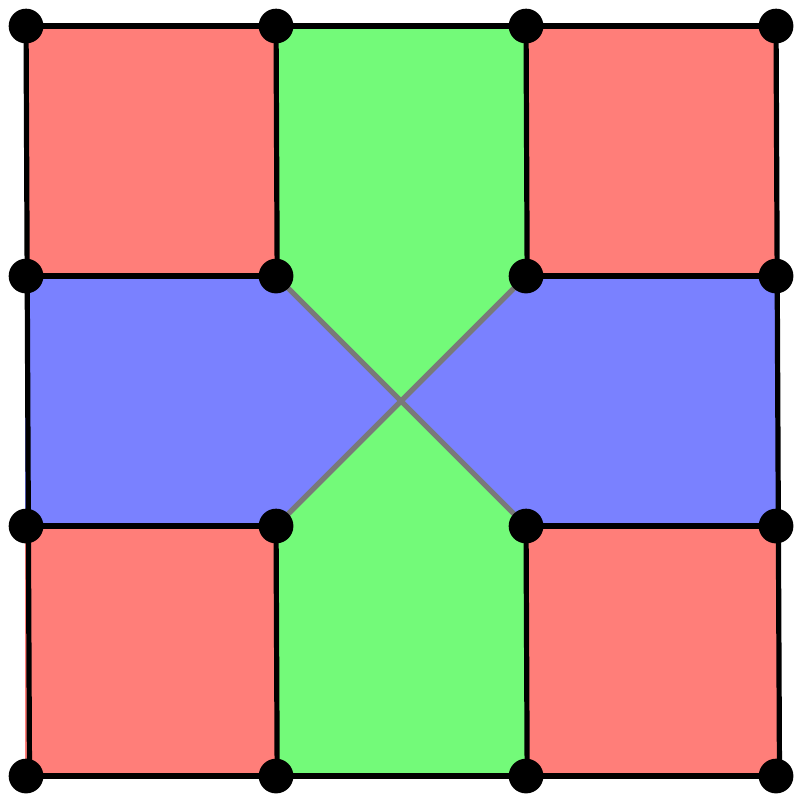}}
\caption{$\llbracket 16,4,4 \rrbracket$ color code.  Two eight-qubit plaquettes consist of those qubits incident to the green and blue regions.  
} \label{f:1644colorcode}
\end{figure}

\begin{figure}
\centering
\begin{align*}
4 \times \; \raisebox{-.85cm}{\includegraphics[scale=.192]{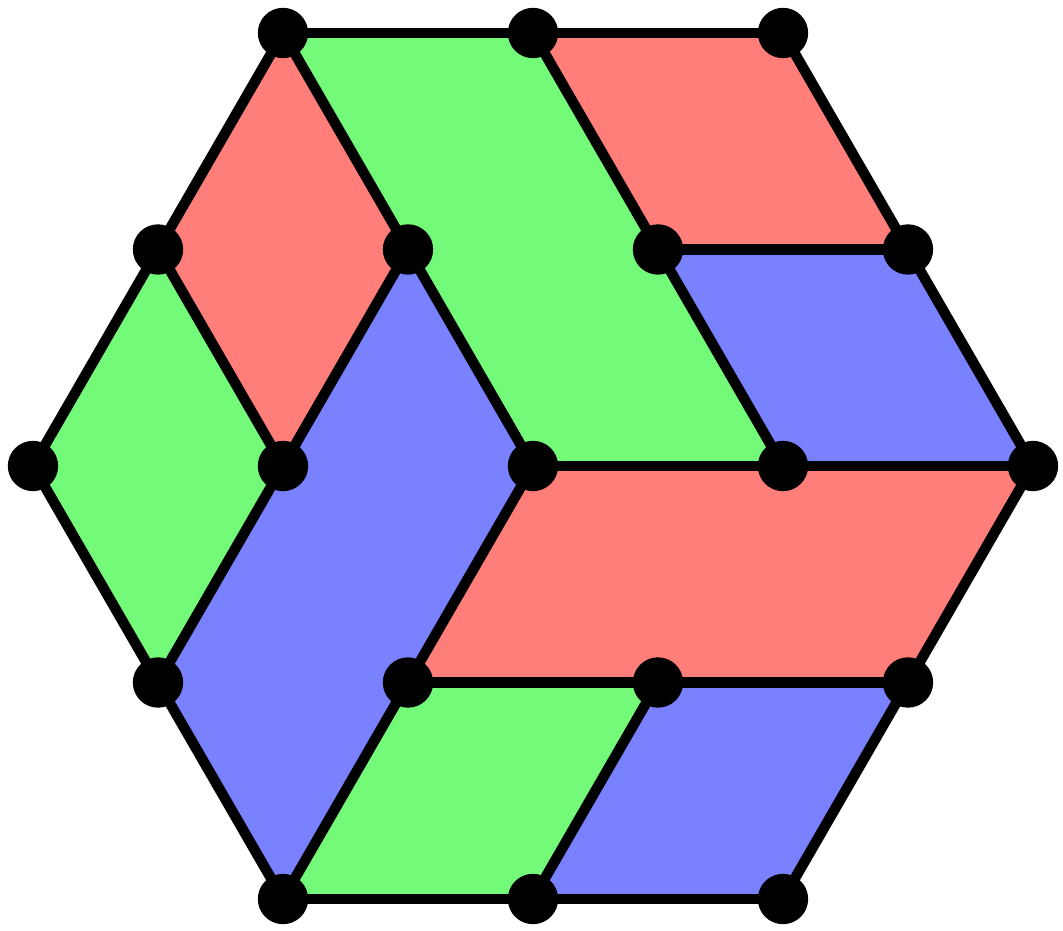}}
\quad \overset{\raisebox{-.85cm}{\includegraphics[scale=.08]{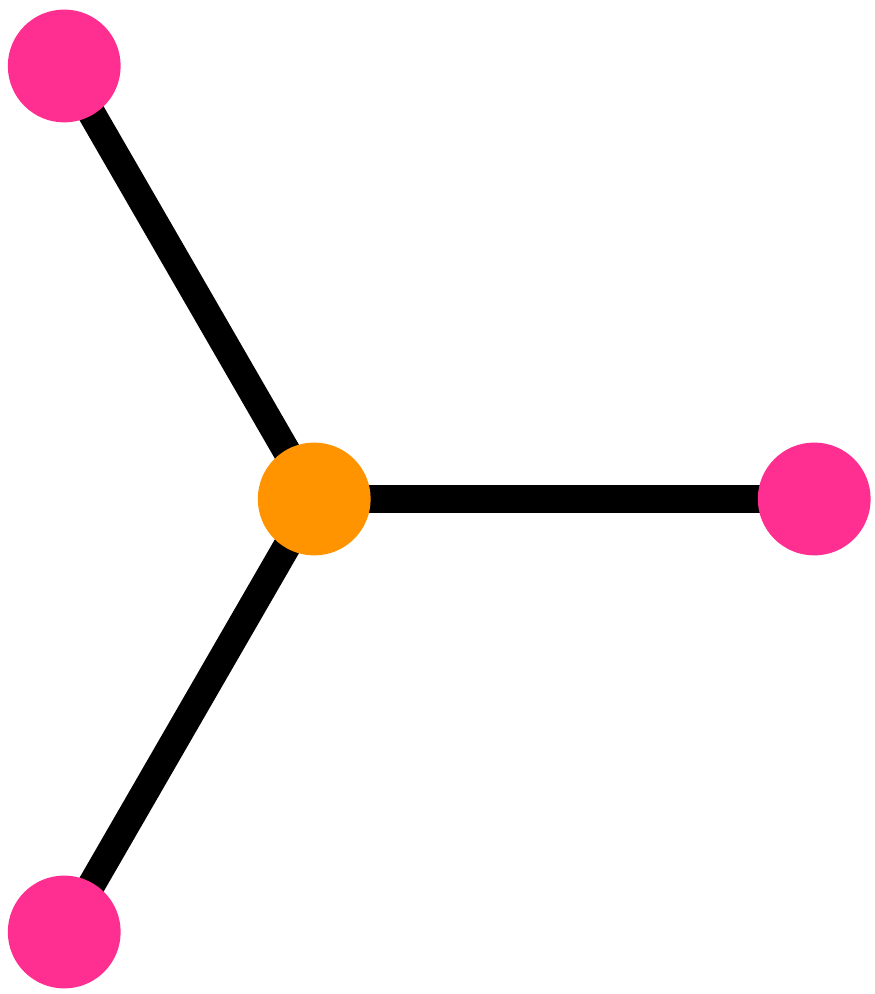}} }{\longrightarrow} \quad 
\raisebox{-2cm}{\includegraphics[scale=.192]{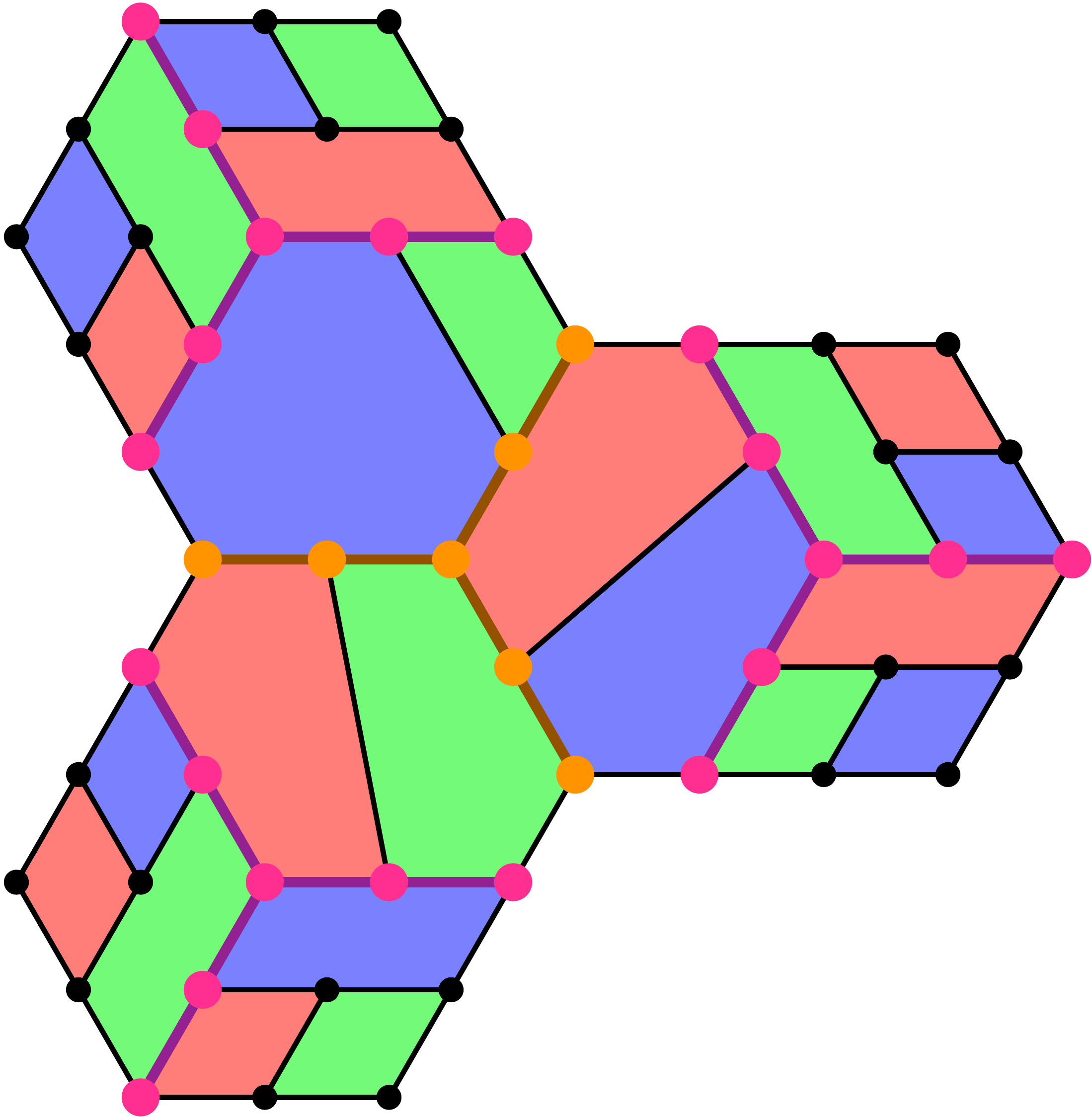}} 
\end{align*}
\caption{
The merged color code construction from \secref{s:mergedcolorcodes} works for higher-distance color codes.  For example, four copies of a $\llbracket 19,1,5 \rrbracket$ color code can be joined along a star graph $K_{3,1}$ to give a $\llbracket 52,4,5 \rrbracket$ color code, shown above.  
We have highlighted the seven-qubit central ``spines" of code blocks for clarity.  
The merge to the upper-left code block is made using four- and eight-vertex plaquettes, and the other merges using six-vertex plaquettes.  
Note that the central encoded qubit is protected to distance six, and in a larger lattice would be protected to distance seven.  
Since the complete graph $K_4$ has a length-three cycle, joining the $\llbracket 19,1,5 \rrbracket$ color code blocks along it will give only a distance-three, $\llbracket 28,4,3 \rrbracket$, code.  
} \label{f:fourdistance5codesmerge}
\end{figure}

Fault-tolerant quantum computation is experimentally accessible even with fairly few qubits.  We have only touched on the design space and areas for optimization.  
We have focused on codes of distance three, capable of correcting a weight-one error.  Chamberland and Beverland~\cite{ChamberlandBeverland17flags} have studied flagged syndrome extraction for the distance-five, $\llbracket 17,1,5 \rrbracket$ and $\llbracket 19,1,5 \rrbracket$ color codes (\figref{f:1715colorcode}).  It is an interesting problem whether the parallel syndrome-extraction techniques we have explored here also extend to these higher-distance codes, and how well the techniques cope with geometrical interaction constraints.  

There are also other moderate-sized color codes of higher-distance, such as a $\llbracket 16,4,4 \rrbracket$ color code (\figref{f:1644colorcode}) that might be preferable to the $\llbracket 16,4,3 \rrbracket$ color codes considered in Sections~\ref{s:onequbitcolorcode} and~\ref{s:mergedcolorcodes}.  
The idea of merging color codes from \secref{s:mergedcolorcodes} generalizes naturally to larger codes (\figref{f:fourdistance5codesmerge}).  
As in Secs.~\ref{s:steanesteaneerrorcorrection} and~\ref{s:codecodeerrorcorrection}, can fault-tolerant error correction on two $\llbracket 19,1,5 \rrbracket$ color code blocks be implemented without extra qubits?

\medskip

Research supported by NSF grant CCF-1254119 and ARO grant W911NF-12-1-0541.

\bibliography{q}

\begin{thebibliography}{18}%
\makeatletter
\providecommand \@ifxundefined [1]{%
 \@ifx{#1\undefined}
}%
\providecommand \@ifnum [1]{%
 \ifnum #1\expandafter \@firstoftwo
 \else \expandafter \@secondoftwo
 \fi
}%
\providecommand \@ifx [1]{%
 \ifx #1\expandafter \@firstoftwo
 \else \expandafter \@secondoftwo
 \fi
}%
\providecommand \natexlab [1]{#1}%
\providecommand \enquote  [1]{``#1''}%
\providecommand \bibnamefont  [1]{#1}%
\providecommand \bibfnamefont [1]{#1}%
\providecommand \citenamefont [1]{#1}%
\providecommand \href@noop [0]{\@secondoftwo}%
\providecommand \href [0]{\begingroup \@sanitize@url \@href}%
\providecommand \@href[1]{\@@startlink{#1}\@@href}%
\providecommand \@@href[1]{\endgroup#1\@@endlink}%
\providecommand \@sanitize@url [0]{\catcode `\\12\catcode `\$12\catcode
  `\&12\catcode `\#12\catcode `\^12\catcode `\_12\catcode `\%12\relax}%
\providecommand \@@startlink[1]{}%
\providecommand \@@endlink[0]{}%
\providecommand \url  [0]{\begingroup\@sanitize@url \@url }%
\providecommand \@url [1]{\endgroup\@href {#1}{\urlprefix }}%
\providecommand \urlprefix  [0]{URL }%
\providecommand \Eprint [0]{\href }%
\providecommand \doibase [0]{http://dx.doi.org/}%
\providecommand \selectlanguage [0]{\@gobble}%
\providecommand \bibinfo  [0]{\@secondoftwo}%
\providecommand \bibfield  [0]{\@secondoftwo}%
\providecommand \translation [1]{[#1]}%
\providecommand \BibitemOpen [0]{}%
\providecommand \bibitemStop [0]{}%
\providecommand \bibitemNoStop [0]{.\EOS\space}%
\providecommand \EOS [0]{\spacefactor3000\relax}%
\providecommand \BibitemShut  [1]{\csname bibitem#1\endcsname}%
\let\auto@bib@innerbib\@empty
\bibitem [{\citenamefont {Steane}(1996)}]{Steane96css}%
  \BibitemOpen
  \bibfield  {author} {\bibinfo {author} {\bibfnamefont {Andrew~M.}\
  \bibnamefont {Steane}},\ }\bibfield  {title} {\enquote {\bibinfo {title}
  {Error correcting codes in quantum theory},}\ }\href {\doibase 10.1103/PhysRevLett.77.793} {\bibfield  {journal} {\bibinfo  {journal} {Phys.
  Rev. Lett.}\ }\textbf {\bibinfo {volume} {77}},\ \bibinfo {pages} {793--797}
  (\bibinfo {year} {1996})}\BibitemShut {NoStop}%
\bibitem [{\citenamefont {Bomb{\'\i}n}\ and\ \citenamefont
  {Martin-Delgado}(2006)}]{BombinMartindelgado06colorcode}%
  \BibitemOpen
  \bibfield  {author} {\bibinfo {author} {\bibfnamefont {H{\'e}ctor}\
  \bibnamefont {Bomb{\'\i}n}}\ and\ \bibinfo {author} {\bibfnamefont
  {Miguel~Angel}\ \bibnamefont {Martin-Delgado}},\ }\bibfield  {title}
  {\enquote {\bibinfo {title} {Topological quantum distillation},}\ }\href
  {\doibase 10.1103/PhysRevLett.97.180501} {\bibfield  {journal} {\bibinfo
  {journal} {Phys. Rev. Lett.}\ }\textbf {\bibinfo {volume} {97}},\ \bibinfo
  {pages} {180501} (\bibinfo {year} {2006})},\ \Eprint
  {http://arxiv.org/abs/quant-ph/0605138} {arXiv:quant-ph/0605138}
  \BibitemShut {NoStop}%
\bibitem [{\citenamefont {Steane}(1997)}]{Steane97}%
  \BibitemOpen
  \bibfield  {author} {\bibinfo {author} {\bibfnamefont {Andrew~M.}\
  \bibnamefont {Steane}},\ }\bibfield  {title} {\enquote {\bibinfo {title}
  {Active stabilization, quantum computation, and quantum state synthesis},}\
  }\href {\doibase 10.1103/PhysRevLett.78.2252} {\bibfield  {journal} {\bibinfo
   {journal} {Phys. Rev. Lett.}\ }\textbf {\bibinfo {volume} {78}},\ \bibinfo
  {pages} {2252--2255} (\bibinfo {year} {1997})},\ \Eprint
  {http://arxiv.org/abs/quant-ph/9611027} {arXiv:quant-ph/9611027}
  \BibitemShut {NoStop}%
\bibitem [{\citenamefont {Steane}(2002)}]{Steane02}%
  \BibitemOpen
  \bibfield  {author} {\bibinfo {author} {\bibfnamefont {Andrew~M.}\
  \bibnamefont {Steane}},\ }\href@noop {} {\enquote {\bibinfo {title} {Fast
  fault-tolerant filtering of quantum codewords},}\ } (\bibinfo {year}
  {2002}),\ \Eprint {http://arxiv.org/abs/quant-ph/0202036}
  {arXiv:quant-ph/0202036} \BibitemShut {NoStop}%
\bibitem [{\citenamefont {Shor}(1996)}]{Shor96}%
  \BibitemOpen
  \bibfield  {author} {\bibinfo {author} {\bibfnamefont {Peter~W.}\
  \bibnamefont {Shor}},\ }\bibfield  {title} {\enquote {\bibinfo {title}
  {Fault-tolerant quantum computation},}\ }in\ \href {\doibase 10.1109/SFCS.1996.548464} {\emph {\bibinfo {booktitle} {Proc. 37th Symp. on
  Foundations of Computer Science (FOCS)}}}\ (\bibinfo {year} {1996})\
  p.~\bibinfo {pages} {96},\ \Eprint
  {http://arxiv.org/abs/quant-ph/9605011} {arXiv:quant-ph/9605011}
  \BibitemShut {NoStop}%
\bibitem [{\citenamefont {DiVincenzo}\ and\ \citenamefont
  {Aliferis}(2007)}]{DiVincenzoAliferis06slow}%
  \BibitemOpen
  \bibfield  {author} {\bibinfo {author} {\bibfnamefont {David~P.}\
  \bibnamefont {DiVincenzo}}\ and\ \bibinfo {author} {\bibfnamefont {Panos}\
  \bibnamefont {Aliferis}},\ }\bibfield  {title} {\enquote {\bibinfo {title}
  {Effective fault-tolerant quantum computation with slow measurements},}\
  }\href {\doibase 10.1103/PhysRevLett.98.020501} {\bibfield  {journal}
  {\bibinfo  {journal} {Phys. Rev. Lett.}\ }\textbf {\bibinfo {volume} {98}},\
  \bibinfo {pages} {220501} (\bibinfo {year} {2007})},\ \Eprint
  {http://arxiv.org/abs/quant-ph/0607047} {arXiv:quant-ph/0607047}
  \BibitemShut {NoStop}%
\bibitem [{\citenamefont {Yoder}\ and\ \citenamefont
  {Kim}(2017)}]{YoderKim16trianglecodes}%
  \BibitemOpen
  \bibfield  {author} {\bibinfo {author} {\bibfnamefont {Theodore~J.}\
  \bibnamefont {Yoder}}\ and\ \bibinfo {author} {\bibfnamefont {Isaac~H.}\
  \bibnamefont {Kim}},\ }\bibfield  {title} {\enquote {\bibinfo {title} {The
  surface code with a twist},}\ }\href {\doibase 10.22331/q-2017-04-25-2}
  {\bibfield  {journal} {\bibinfo  {journal} {Quantum}\ }\textbf {\bibinfo
  {volume} {1}},\ \bibinfo {pages} {2} (\bibinfo {year} {2017})},\ \Eprint
  {http://arxiv.org/abs/1612.04795} {arXiv:1612.04795
  [quant-ph]} \BibitemShut {NoStop}%
\bibitem [{\citenamefont {Chao}\ and\ \citenamefont
  {Reichardt}(2017)}]{ChaoReichardt17errorcorrection}%
  \BibitemOpen
  \bibfield  {author} {\bibinfo {author} {\bibfnamefont {Rui}\ \bibnamefont
  {Chao}}\ and\ \bibinfo {author} {\bibfnamefont {Ben~W.}\ \bibnamefont
  {Reichardt}},\ }\bibfield  {title} {\enquote {\bibinfo {title} {Error
  correction with only two extra qubits},}\ }\href@noop {} {\  (\bibinfo {year}
  {2017})},\ \Eprint {http://arxiv.org/abs/1705.02329}
  {arXiv:1705.02329 [quant-ph]} \BibitemShut {NoStop}%
\bibitem [{\citenamefont {Chamberland}\ and\ \citenamefont
  {Beverland}(2018)}]{ChamberlandBeverland17flags}%
  \BibitemOpen
  \bibfield  {author} {\bibinfo {author} {\bibfnamefont {Christopher}\
  \bibnamefont {Chamberland}}\ and\ \bibinfo {author} {\bibfnamefont
  {Michael~E.}\ \bibnamefont {Beverland}},\ }\bibfield  {title} {\enquote
  {\bibinfo {title} {Flag fault-tolerant error correction with arbitrary
  distance codes},}\ }\href {\doibase 10.22331/q-2018-02-08-53} {\bibfield
  {journal} {\bibinfo  {journal} {Quantum}\ }\textbf {\bibinfo {volume} {2}},\
  \bibinfo {pages} {53} (\bibinfo {year} {2018})},\ \Eprint
  {http://arxiv.org/abs/1708.02246} {arXiv:1708.02246
  [quant-ph]} \BibitemShut {NoStop}%
\bibitem [{\citenamefont {Tansuwannont}\ \emph {et~al.}(2018)\citenamefont
  {Tansuwannont}, \citenamefont {Chamberland},\ and\ \citenamefont
  {Leung}}]{TansuwannontChamberlandLeung18flag}%
  \BibitemOpen
  \bibfield  {author} {\bibinfo {author} {\bibfnamefont {Theerapat}\
  \bibnamefont {Tansuwannont}}, \bibinfo {author} {\bibfnamefont {Christopher}\
  \bibnamefont {Chamberland}}, \ and\ \bibinfo {author} {\bibfnamefont
  {Debbie}\ \bibnamefont {Leung}},\ }\href@noop {} {\enquote {\bibinfo {title}
  {Flag fault-tolerant error correction for cyclic {CSS} codes},}\ } (\bibinfo
  {year} {2018}),\ \Eprint {http://arxiv.org/abs/1803.09758}
  {arXiv:1803.09758 [quant-ph]} \BibitemShut {NoStop}%
\bibitem [{\citenamefont {Barends}\ \emph {et~al.}(2014)\citenamefont
  {Barends}, \citenamefont {Kelly}, \citenamefont {Megrant}, \citenamefont
  {Veitia}, \citenamefont {Sank}, \citenamefont {Jeffrey}, \citenamefont
  {White}, \citenamefont {Mutus}, \citenamefont {Fowler}, \citenamefont
  {Campbell}, \citenamefont {Chen}, \citenamefont {Chen}, \citenamefont
  {Chiaro}, \citenamefont {Dunsworth}, \citenamefont {Neill}, \citenamefont
  {P.~O'Malley}, \citenamefont {Vainsencher}, \citenamefont {Wenner},
  \citenamefont {Korotkov}, \citenamefont {Cleland},\ and\ \citenamefont
  {Martinis}}]{Barendsetal14superconducting}%
  \BibitemOpen
  \bibfield  {author} {\bibinfo {author} {\bibfnamefont {R.}~\bibnamefont
  {Barends}} \emph {et~al.},\ }\bibfield  {title} {\enquote {\bibinfo {title}
  {Superconducting quantum circuits at the surface code threshold for fault
  tolerance},}\ }\href {\doibase 10.1038/nature13171} {\bibfield  {journal}
  {\bibinfo  {journal} {Nature}\ }\textbf {\bibinfo {volume} {508}},\ \bibinfo
  {pages} {500--503} (\bibinfo {year} {2014})},\ \Eprint
  {http://arxiv.org/abs/1402.4848} {arXiv:1402.4848
  [quant-ph]} \BibitemShut {NoStop}%
\bibitem [{\citenamefont {Aliferis}\ and\ \citenamefont
  {Cross}(2007)}]{AliferisCross06BaconShorft}%
  \BibitemOpen
  \bibfield  {author} {\bibinfo {author} {\bibfnamefont {Panos}\ \bibnamefont
  {Aliferis}}\ and\ \bibinfo {author} {\bibfnamefont {Andrew~W.}\ \bibnamefont
  {Cross}},\ }\bibfield  {title} {\enquote {\bibinfo {title} {Subsystem fault
  tolerance with the {B}acon-{S}hor code},}\ }\href {\doibase 10.1103/PhysRevLett.98.220502} {\bibfield  {journal} {\bibinfo  {journal}
  {Phys. Rev. Lett.}\ }\textbf {\bibinfo {volume} {98}},\ \bibinfo {pages}
  {220502} (\bibinfo {year} {2007})},\ \Eprint
  {http://arxiv.org/abs/quant-ph/0610063} {arXiv:quant-ph/0610063}
  \BibitemShut {NoStop}%
\bibitem [{\citenamefont {Li}\ \emph {et~al.}(2018)\citenamefont {Li},
  \citenamefont {Miller},\ and\ \citenamefont
  {Brown}}]{LiMillerBrown18baconshor}%
  \BibitemOpen
  \bibfield  {author} {\bibinfo {author} {\bibfnamefont {Muyuan}\ \bibnamefont
  {Li}}, \bibinfo {author} {\bibfnamefont {Daniel}\ \bibnamefont {Miller}}, \
  and\ \bibinfo {author} {\bibfnamefont {Kenneth~R.}\ \bibnamefont {Brown}},\
  }\href@noop {} {\enquote {\bibinfo {title} {Direct measurement of
  {B}acon-{S}hor code stabilizers},}\ } (\bibinfo {year} {2018}),\ \Eprint
  {http://arxiv.org/abs/1804.01127} {arXiv:1804.01127
  [quant-ph]} \BibitemShut {NoStop}%
\bibitem [{\citenamefont {Knill}(2005{\natexlab{a}})}]{Knill03erasure}%
  \BibitemOpen
  \bibfield  {author} {\bibinfo {author} {\bibfnamefont {Emanuel}\ \bibnamefont
  {Knill}},\ }\bibfield  {title} {\enquote {\bibinfo {title} {Scalable quantum
  computing in the presence of large detected-error rates},}\ }\href {\doibase 10.1103/PhysRevA.71.042322} {\bibfield  {journal} {\bibinfo  {journal} {Phys.
  Rev. A}\ }\textbf {\bibinfo {volume} {71}},\ \bibinfo {pages} {042322}
  (\bibinfo {year} {2005}{\natexlab{a}})},\ \Eprint
  {http://arxiv.org/abs/quant-ph/0312190} {arXiv:quant-ph/0312190}
  \BibitemShut {NoStop}%
\bibitem [{\citenamefont {Aliferis}\ \emph {et~al.}(2006)\citenamefont
  {Aliferis}, \citenamefont {Gottesman},\ and\ \citenamefont
  {Preskill}}]{AliferisGottesmanPreskill05}%
  \BibitemOpen
  \bibfield  {author} {\bibinfo {author} {\bibfnamefont {Panos}\ \bibnamefont
  {Aliferis}}, \bibinfo {author} {\bibfnamefont {Daniel}\ \bibnamefont
  {Gottesman}}, \ and\ \bibinfo {author} {\bibfnamefont {John}\ \bibnamefont
  {Preskill}},\ }\bibfield  {title} {\enquote {\bibinfo {title} {Quantum
  accuracy threshold for concatenated distance-3 codes},}\ } {\bibfield  {journal} {\bibinfo
  {journal} {Quant. Inf. Comput.}\ }\textbf {\bibinfo {volume} {6}},\ \bibinfo
  {pages} {97--165} (\bibinfo {year} {2006})},\ \Eprint
  {http://arxiv.org/abs/quant-ph/0504218} {arXiv:quant-ph/0504218} \BibitemShut
  {NoStop}%
\bibitem [{\citenamefont {Knill}(2005{\natexlab{b}})}]{Knill05}%
  \BibitemOpen
  \bibfield  {author} {\bibinfo {author} {\bibfnamefont {Emanuel}\ \bibnamefont
  {Knill}},\ }\bibfield  {title} {\enquote {\bibinfo {title} {Quantum computing
  with realistically noisy devices},}\ }\href {\doibase 10.1038/nature03350}
  {\bibfield  {journal} {\bibinfo  {journal} {Nature}\ }\textbf {\bibinfo
  {volume} {434}},\ \bibinfo {pages} {39--44} (\bibinfo {year}
  {2005}{\natexlab{b}})}\BibitemShut {NoStop}%
\bibitem [{\citenamefont {Tomita}\ and\ \citenamefont
  {Svore}(2014)}]{TomitaSvore14surfacecode}%
  \BibitemOpen
  \bibfield  {author} {\bibinfo {author} {\bibfnamefont {Yu}~\bibnamefont
  {Tomita}}\ and\ \bibinfo {author} {\bibfnamefont {Krysta~M.}\ \bibnamefont
  {Svore}},\ }\bibfield  {title} {\enquote {\bibinfo {title} {Low-distance
  surface codes under realistic quantum noise},}\ }\href {\doibase 10.1103/PhysRevA.90.062320} {\bibfield  {journal} {\bibinfo  {journal} {Phys.
  Rev. A}\ }\textbf {\bibinfo {volume} {90}},\ \bibinfo {pages} {062320}
  (\bibinfo {year} {2014})},\ \Eprint {http://arxiv.org/abs/1404.3747} {arXiv:1404.3747 [quant-ph]} \BibitemShut {NoStop}%
\bibitem [{\citenamefont {Bacon}(2006)}]{Bacon05operator}%
  \BibitemOpen
  \bibfield  {author} {\bibinfo {author} {\bibfnamefont {Dave}\ \bibnamefont
  {Bacon}},\ }\bibfield  {title} {\enquote {\bibinfo {title} {Operator quantum
  error correcting subsystems for self-correcting quantum memories},}\ }\href
  {\doibase 10.1103/PhysRevA.73.012340} {\bibfield  {journal} {\bibinfo
  {journal} {Phys. Rev. A}\ }\textbf {\bibinfo {volume} {73}},\ \bibinfo
  {pages} {012340} (\bibinfo {year} {2006})},\ \Eprint
  {http://arxiv.org/abs/quant-ph/0506023} {arXiv:quant-ph/0506023}
  \BibitemShut {NoStop}%
\end{thebibliography}%

\end{document}